\documentclass[aps,preprint,epsf,epsfig,amsmath]{revtex4}
\usepackage[toc,page]{appendix}
\usepackage{graphicx}
\usepackage{alltt,dsfont,amsmath,bm}

\usepackage{color}

\newcommand{\col}{\color{black}}

\newcommand{\G}{{\cal{G}}}

\newcommand{\GH}{{\bf g}}

\newcommand{\si}{\sigma}

\newcommand{\tJ}{\ $t$-$J$ \ }

\newcommand{\nn}{\nonumber}

\newcommand{\beq}{\begin{equation}}
\newcommand{\eeq}{\end{equation}}
\newcommand{\barray}{\begin{eqnarray}}
\newcommand{\earray}{\end{eqnarray}}
\newcommand{\disp}[1]{Eq.~(\ref{#1})}
\newcommand{\refdisp}[1]{Ref.~(\onlinecite{#1})}
\newcommand{\figdisp}[1]{Fig.~(\ref{#1})}

\newcommand{\g}{\bar{G}}
\newcommand{\E}{E^{ FL}}
\newcommand{\xxi}{\varepsilon_0}
\begin{document}
\title{ Anatomy of the Self Energy\\ }
\author{ B. Sriram Shastry}
\affiliation{ Physics Department, University of California, Santa Cruz, CA 95064, USA}
\date{October 11, 2011}

\begin{abstract}

{\col  The general problem of  representing the  Greens function $G(k,z)$ in terms of self energy, in  field theories    lacking Wick's theorem,   is considered.  A simple  construction shows that a Dyson like representation with a  self energy $\Sigma(k,z)$ is always possible, provided  we start with a spectral representation for $G(k,z)$ for finite  systems and take the thermodynamic limit. The self energy itself can  then be iteratively expressed in terms of another higher order self energy,  capturing the spirit of Mori's formulation.}

{\col We further discuss alternative and  more general forms  of $G(k,z)$ that are  possible. In particular,  }
a recent theory {\col by the author} of extremely correlated Fermi liquids at density ``$n$",  for  Gutzwiller projected non canonical Fermi operators obtains a new form of the Greens function: $$G(k,z)= \frac{ \left\{ 1- \frac{n}{2}\right\} + \Psi(k,z)}{z - \hat{E}_k - \Phi(k,z)}, $$  with {\em  a pair of self energies} $\Phi(z)$ and $\Psi(z)$.  Its  relationship with the Dyson form is explored.  A simple   version of the two self energy model  {\col   was shown recently to successfully fit several  data sets of photoemission line shapes in cuprates.}   We provide {\col details of } the  unusual spectral line shapes  that arise in this model, with the characteristic skewed shape depending  upon a single parameter.  The EDC and MDC line shapes are shown to be skewed in opposite directions, and provide  a testable prediction of the theory.
 
 \end{abstract}

\maketitle

\date{\today}
\maketitle
\section{Introduction}

{\col    Our  work  
explores  the    representation of the  Greens function $G(k,z)$ of a particle, in  field theories    without   Wick's theorem. While Wick's theorem for  Bosons and  Fermions automatically gives rise to the Dyson form of self energy,  its absence   for non canonical, i.e. general operators (other than Bosons of Fermions),  leads to a conundrum that is poorly understood. This work addresses a particular type on non canonical theory originating from Gutzwiller projection of electrons on a lattice.}

The  representation of the propagator or the Greens function  in terms of its  Dyson self energy is a fundamental paradigm of standard interacting relativistic and non relativistic field theories. The structure of this representation and  the generation of approximate approximations for the self energy in terms of  Feynman diagrams, the vertex function or higher order Greens functions form  the dominant part of  existing literature of many particle physics. 

 In the context of extending these studies to  extremely large and singular repulsive  interactions, termed extreme correlations\cite{ECFL},  one needs to deal with non canonical electrons. A standard non-canonical problem involves    the Hubbard operators\cite{ECQL} $X_j^{a, b}$ located at  sites $j$ of a lattice.  These are ``graded''  projection operators  with label  $a$ representing the three allowed local configurations $0, \uparrow, \downarrow$.   Among these operators, $X_j^{0 \si}, X_j^{\si 0}$ are Fermionic  destruction and creation type  objects.  Their Greens function is  measured directly in  angle resolved photoemission experiments (ARPES) on certain  experimental systems embodying extreme correlations,  including the high temperature superconductors\cite{gweon-ecfl}. 

Quite recently the author has formulated   in  \refdisp{ECFL},  the theory of an  Extremely Correlated Fermi Liquid (ECFL) state of the $t$-$J$ model,    where he  has  found another type of representation for the Greens function with {\em a pair of self energies} (see \disp{twog2} below)  by using the non perturbative Schwinger approach of source fields to depict the  equations of motion. The use of more general forms of Greens functions is not completely new, there are examples in literature of multiple self energies in  \refdisp{logan} and \refdisp{kotliar}. 
  The physics of extreme correlations treated here, is based on non perturbative considerations without obvious parallels in  weak or intermediate coupling problems. It  leads to the two self energy form \disp{twog2}, whose distinctive signatures  are strikingly different from those of Fermi liquids.
 
 The technical details of the  construction in \refdisp{ECFL} are intricate and require the processing of two parallel hierarchies for the two self energies.  {\col A separate paper is in preparation detailing the  involved technical details and the calculation in \refdisp{ECFL}.}  {\col    Further  background details of the notation, definitions and sum rules satisfied by   the Greens functions  for extreme correlations, and its analyticity can be found in the earlier publication  \refdisp{ECQL}. } A suggestive  functional form of the Greens function \disp{spectral-8} emerges  from \refdisp{ECFL},  by making the assumption of momentum independence of the two self energies, valid  in high dimensions.  It  satisfies the number sum rule and the total  particle weight integrates to unity in each state.   In  \refdisp{gweon-ecfl},  Gweon, Gu  and the author have shown that  several experimental data sets  on ARPES by different groups using both the traditional synchrotron source light and laser source light can be reconciled very well  with the line shape  in \disp{spectral-8} and \disp{peak-3}. This is {\em the first satisfactory  functional  form that has  been found to fit both laser and synchrotron data}, and works very well with  few adjustable parameters.
 
    Therefore a  major objective of this paper is to elucidate  the detailed form of the spectral lines that emerge from the above  {\col simple version of the } ECFL theory in \refdisp{ECFL} and successfully employed to understand experiments in \refdisp{gweon-ecfl}. 
  {\col Our hope is that this detailed analysis would familiarize readers  with the nuances of the new spectral function, and thereby  facilitate  ARPES line shape analysis  of  further experiments on high temperature superconductors and other materials, in a manner analogous to that in \refdisp{gweon-ecfl}. }
 
  For the above purpose, we recall that in a lattice of finite (say small)  number of sites,  the state space is finite dimensional and hence the Greens function for arbitrarily complicated objects can be computed  by numerical means,   leading to  rational  functions of the complex frequency $z$  as in  \disp{eq-g} below.  We begin by studying this representation and see how the Dyson representation arises;  we find that the two self energy representation \disp{twog2} is also quite  natural from this view point.
We further study the infinite size limits where the poles coalesce to give cuts in the complex $z$ plane. 
We adopt a phenomenological model for an underlying auxiliary  Fermi Liquid (aux-FL) self energy, enabling us to display  analytic expressions  for the   Greens function. We provide a detailed perspective on the representation in \disp{spectral-8}, namely the location of the poles and the subtle differences from a standard Fermi liquid.

Another  result in this paper is to show that a Dyson like representation with a  self energy $\Sigma(k,z)$ is always possible, provided  we start with a spectral representation. The self energy itself can  then be iteratively expressed in terms of another higher order self energy. 
 This hierarchical result is cast in the same form  as   the Mori formalism. While the Mori formalism is very abstract, and expressed in terms of projection operators, we can    go beyond   it in a certain sense. By  working with standard spectral representation,  we show that  it is  possible to express the  higher self energy spectral functions in terms of the lower ones, leading to  an explicit hierarchy. Our construction completely bypasses the Mori projection operators, and should be useful in throwing light on the latter.

The plan of the paper is as follows. In Sec (II) and (III) we note the spectral representation and study the 
Greens function as a rational function of complex frequency $z$ for a finite system. In Sec (IV) we note the representation in the limit of infinite size and introduce the high dimensional expression with two self energies. The detailed structure of the characteristic  line shape as in \disp{peak-3} is discussed, and its dependence on physical parameters displayed with the help of a numerical example.  An explicit example of the auxiliary Fermi liquid Greens function is provided, and typical values of the parameters are argued for. In Sec (V) the line shapes in EDC and MDC are displayed in detail, in order to bring out the specific signatures of the theory, namely a skew in the spectrum arising from the caparison factor in \refdisp{ECFL} and \disp{spectral-8}. 
In Sec(VI) we discuss the amusing connection with higher order self energies of the type that Mori's formalism yields, but at a much more explicit level than what is available in literature.

\section{spectral representation of the Greens function}
Let us begin with the spectral representation\cite{agd} of the Matsubara Greens function at finite temperatures given by: 
\beq
G(k, z) =  \int \ dx \ \frac{\rho_G(k,x)}{z-x}, \label{eq1}
\eeq
where $G$ is the Greens function at a fixed wave vector $k$, and $\rho_G$ is its spectral density and the integration range is   $- \infty \leq x \leq \infty$. 
To simplify notation, we call the Greens function as $G(k,z)$,  the same object was  denoted by $\G(k,z)$ in \refdisp{ECFL}.
 The index $k$ can be also replaced by a spatial index when dealing with a local Greens function.  The spectral function $\rho_G(k,x)$  in most problems of interest  in Condensed Matter Physics has a compact support, so that $G(k,z)$ has   ``reasonable'' behaviour in the complex $z$ plane, with an asymptotic $1/z$ fall off,  and apart from  a branch cut on a portion of the real line, is analytic. The frequency $z$ is either Fermionic or Bosonic depending on the statistics of the underlying particles. The  spectral function is given by the standard formula\cite{agd,ECQL}
\beq
\rho_G(k,x)= \sum_{\alpha, \beta} | \langle \alpha | A(k) | \beta \rangle |^2 (p_\alpha + p_\beta) \ \delta(x+ \varepsilon_\alpha-\varepsilon_\beta), \label{eq2}
\eeq
where $A(k)$ is the destruction operator, $p_\alpha $ is the Boltzmann probability of the state $\alpha$ given by $e^{- \beta \varepsilon_\alpha}/Z$ and $\varepsilon_\alpha$ is the eigenvalue of the grand  Hamiltonian of the system $K= H- \mu \hat{N}$.  In the case of canonical particles $A(k)$ is the usual Fermi or Bose destruction operator.  In \refdisp{ECFL},    non canonical  Hubbard ``X'' operators are  considered;  we will not require any detailed information about them here except that the anticommutator $ \left\{A,A^\dagger\right\} $ is not unity,  but rather an object with a known expectation value ~($ 1-n/2$), in terms of   the dimensionless particle density  $n$.

We  consider two alternate representations of the Greens function in terms of the complex frequency $z$  that are available in many-body physics: (a) for canonical Bosons or Fermions  the  Dyson  representation in terms of a single self energy $\Sigma(z)$ and (b)  for non canonical particles a novel form proposed recently by the author  with {\em two} self energy type objects $\Phi(z)$ and $\Psi(z)$. 
\barray
G(k,z) &= & \frac{a_G}{z - \hat{E}_k - \Sigma(k,z) },\;\;\;\;\mbox{Dyson}\label{twog1}\\
&=& \frac{a_G + \Psi(k,z)}{z - \hat{E}_k - \Phi(k,z)}. \;\;\;\;\mbox{ECFL}\label{twog2}
\earray
For canonical objects $a_G=1$  and for Hubbard operators in the ECFL we write $a_G= 1- n/2$.
 We  start below  from a finite  system, where the Greens function is a meromorphic function expressible as the  sum over isolated poles in the complex frequency plane with given residues. 
 In fact it is a rational function as well, expressible as the ratio of two polynomials.   Using  simple arguments,  we will see that 
the above two representations in \disp{twog1} and \disp{twog2} are both natural ways of proceeding with the self energy concept.
 In the limit of a large system, the poles coalesce to give us cuts in the complex frequency plane with specific  spectral densities. In this limit  we display the equations relating the different spectral functions.

\section{Finite system Greens function}

We  drop the explicit mention of the wave vector k, and start with the case of a finite-size system, where we may diagonalize the system exactly and assemble the Greens function from the matrix elements of the operators $A$ and the eigenenergies as in \disp{eq2}.  We see that $\rho_G$ is a sum over say $m$ delta functions located at the eigenenergies $E_j$ (assumed distinct), so we can write the meromorphic representation
\beq
\g(z) = \sum_{j=1}^{m} \frac{a_j}{z- E_j}. \label{eq-g}
\eeq
 The overbar in $\g(z)$ is to emphasize that we are dealing with the finite-size version of the Greens function $G(z)$. Here  $a_j, E_j$ constitute $2 m$ known real parameters.   The sum
\beq
\sum_{j=1}^m a_j = a_{G}, \label{eq-asum}
\eeq
where $a_{G}=1$ for canonical objects and we denote  $a_{G} =1-n/2$ for the   non canonical case of ECFL. In the infinite size limit we set $\g(z)  \to G(z)$.  It is clear that for $z \gg \{ E_j \}_{max}$ we get the asymptotic behaviour $\g\to \frac{a_G}{z}$,  and therefore $\g$ is a rational function that may be expressed as the ratio of two polynomials in $z$ of degrees $m-1$ and $m$:
\beq
\g(z) = a_G \frac{P_{m-1}(z)}{Q_{m}(z)}, \;\;\; Q(z) = \prod_{j=1}^m (z- E_j),\;\;\;P(z)=  \prod_{r=1}^{m-1} (z- \gamma_r), \label{polynomial-1}
\eeq
where the roots $\gamma_r$  are expressible in terms of $a_j$  \& $E_j$. We use the convention that all polynomials  $Q_m \ldots$ have the  coefficient of the leading power of $z$ as unity,   and the degree is indicated explicitly.

We now proceed to find the self energy type expansion for $\g$, and for this purpose
multiplying \disp{eq-g} by $z$ and rearranging we get the ``equation of motion'':
\beq
(z- \hat{E}) \g(z) = a_{G} + \bar{I}(z), \label{eq-eom1}
\eeq
where we introduced a mean energy $\hat{E}$
\barray
\hat{E}&=& \frac{1}{a_{G}}  \sum a_j E_j \nn \\
\bar{I}(z)&=& \sum_{j=1}^{m}  \frac{a_j (E_j-\hat{E})}{z- E_j}, \label{eq-I}
\earray
so that  asymptotically at large $z$, we get $\bar{I}(z) \sim O(1/z^2)$. In standard theory $\hat{E}$ plays the role of the Hartree Fock self energy so that the remaining self energy vanishes at high frequencies  \cite{fn2} . Motivated by the structure of the theory of extremely correlated Fermi systems \refdisp{ECFL}, we next introduce the basic decomposition:
\beq
\bar{I}(z) = \g(z) \Phi(z) + \Psi(z), \label{eq-I2}
\eeq 
where we have introduced two  self energy type functions $\Phi(z)$ and $\Psi(z)$ that will be determined next.  Clearly \disp{eq-I2} leads immediately to the Greens function \disp{twog2} (or \disp{twog1} if we set $\Psi\to 0$). The rationale for \disp{eq-I2} lies in the fact that the function $\bar{I}$ has  the same poles as $\g(z)$. Thus  it has a representation as a ratio of two polynomials
\beq
\bar{I}(z) = i_0 \frac{R_{m-2}(z)}{Q_m(z)}, \label{polynomial-2}
\eeq 
with $R_{m-2}$ a polynomial of degree $m-2$ and $i_0$ a suitable constant and {\em the same polynomial $Q$ from } \disp{polynomial-1}, thereby it is natural to seek a proportionality with $\g$ itself.
If we drop $\Psi$ and rename $\Phi\to \Sigma$, then this gives the usual Dyson self energy $\Sigma(z)$ determined uniquely  using \disp{polynomial-1} and \disp{polynomial-2} as:
\beq
\Sigma(z)= \frac{i_0}{a_G} \frac{R_{m-2}(z)}{P_{m-1}(z)}. \label{polynomial-3}
\eeq

The expression \disp{eq-I2} offers a more general possibility, where  $\Phi(z)$ and $\Psi(z)$ may be viewed as the quotient and remainder obtained by dividing 
$\bar{I}(z)$  by $\g(z)$. It is straightforward to see that $\Psi(z)$ and $\Phi(z)$ are also rational functions expressible as ratios of two polynomials 
\beq
\Psi(z)= \psi_o \frac{K_{m-3}(z)}{D_{m-1}(z)},\;\;\; \Phi(z)= \phi_o \frac{L_{m-2}(z)}{D_{m-1}(z)}, \label{polynomial-33}
\eeq
where $K,L,D$  are polynomials of the displayed degree. Comparing the poles and the zeros  of $\g$ beween \disp{twog2} with \disp{polynomial-3} and \disp{polynomial-1}, we  write down two equations
\barray
a_G P_{m-1}&=& a_G D_{m-1} + \psi_o K_{m-3} \nn \\
Q_m&=& (z- \bar{E}) D_{m-1} - \phi_0 L_{m-2}, \label{polynomial-4}
\earray  
so that we may eliminate $D$ and write an identity
\beq
(z-\bar{E}) P_{m-1}-Q_m = \frac{\psi_o}{a_G} (z-\bar{E})  K_{m-3}+ \phi_o L_{m-2}. \label{polynomial-44}
\eeq
Here the LHS is assumed known and we have two polynomials to determine from this equation. Therefore there are multiple solutions of this problem, and indeed setting $K\to 0$ gives the Dyson form as a special case.

\subsection{A simple   example with two sites:}

The Greens function of the \tJ model at density $n$ with $J=0$ and only two sites is a trivial problem that illustrates the 
two possibilities discussed above.  The two quantum numbers $k=0,\pi$ correspond to the  bonding and antibonding states with energies $e_k= \mp t$, and  a simple calculation at a given $k$ gives \disp{eq-g} as
\beq
\g(k, z)= \frac{a_1 }{z-e_k}+\frac{a_2 }{z+e_k}, \label{size2-1}
\eeq
where $z= i \omega_n + \mu$,  $a_2= e^{\beta \mu}(1+ e^{\beta (\mu-e_k)})/(2Z)$, $a_1= 1-n/2- a_2$ and  the grand partition function $Z= 1+ 4 e^{2 \beta \mu}+ 4 e^{ \beta \mu} \cosh(\beta t)$.  This can be readily expressed as
\beq
\g(k,z)= \frac{(1- \frac{n}{2}) + \Psi(k,z)}{z- E_k - \Phi(k,z)},\;\;\Psi(k,z)= \frac{B_k}{z+E_k},\;\;\Phi(k,z)= \frac{A_k}{z+ E_k}, \label{size2-2}
\eeq
where $E_k$ is arbitrary, $A_k= (E_k^2-e_k^2)$ and  $B_k= (1-n/2)( \bar{E}_k-E_k)$ and with the first moment of energy 
$\bar{E}_k= e_k (a_1-a_2)/(1-n/2)$. As we expected,  the functions $\Psi,\Phi$ thus have a single pole, as opposed to $\g$ with two poles. In this case the dynamics it rather trivial so that the choice of $E_k$ is free. If  we set $E_k=\bar{E}_k$ the residue $B_k$ vanishes and so the second form collapses.

\subsection{Summary of analysis:}
{\col  In summary,  guided by analyticity and the pole structure of $\overline{G}(k,z)$,  we  find it possible to go beyond the standard Dyson representation.   However, we end up getting more freedom than we might have naively expected. This excess freedom is not unnatural, since we haven't yet discussed the microscopic origin of these two self energies.  The theory in \refdisp{ECFL} provides an explicit expression for the two objects $\Psi$ and $\Phi$, where a common linear functional differential operator $\bf{L}$ generates these self energies by acting upon different ``seed" functions as in Eq.~(7) of \refdisp{ECFL}.  The above discussion therefore provides some intuitive understanding of the novel form of the Greens function in  \disp{twog2}, without actually providing an alternative  derivation to that in \refdisp{ECFL}. 
  }

\section{Infinite system Spectral densities and relationships}
In the infinite size limit,  the various functions will be represented in terms of spectral densities obtained from the coalescing of the poles.  Following \disp{eq1}
we will denote a general function 
\beq
Q(z) =  \int \ dx \ \frac{\rho_Q(x)}{z-x}, \label{eq-10}
\eeq
where $Q= \Sigma, \Phi, \Psi$, in terms of its density $\rho_Q(x)$.  The density is given by $\rho_Q(x)= (- \frac{1}{\pi}) \Im m \ Q(x+ i 0^+)$, as usual.
In parallel to the discussion of \disp{eq1}, the assumption of a compact support of $\rho_Q$ gives us well behaved functions. We now turn to the objective of relating the spectral functions in the two representations discussed above.

\subsection{ \bf  Spectral representation for the  Dyson Self Energy}
Let us start with \disp{eq1} and the standard Dyson form \disp{twog1} where we drop the overbar  and study the infinite system function $G(z)$. We use the symbolic identity:
\beq
\frac{1}{x+ i 0^+}= {\cal P}\frac{1}{x} - i \pi \delta(x),
\eeq
 with real $x$,  ${\cal P}$ denoting the principal value, and
where the Hilbert transform of a function $f(u)$ is defined by
\beq
{\cal H}[f](x) = {\cal P} \int_{-\infty}^{\infty}\ d y \ \frac{f(y)}{x-y}.
\eeq
We note  the following  standard result for completeness:
\beq
\rho_G(x) = a_{G} \ \frac{\rho_{\Sigma}(x)}{( \pi \rho_{\Sigma}(x))^2 + \left(x- \hat{E}-  {\cal H}[\rho_\Sigma](x)\right)^2 }. \label{direct}
\eeq
A more  interesting inverse problem is  to solve  for $\rho_\Sigma(x)$ given $G(z)$. Towards this end we rewrite the Dyson equation as
\beq
\Sigma(z)=z- \hat{E} - \frac{a_{G}}{G(z)}, \label{dyson-inverted}
\eeq
where the  self energy   vanishes  asymptotically as $1/z$  provided  the constant part, if any, is absorbed in $\hat{E}$. Therefore  this object can be decomposed in the  fashion of \disp{eq-10}. We compare \disp{dyson-inverted} with \disp{eq-10} with $Q \to \Sigma$ and  conclude that
\beq
\rho_{\Sigma}(x) =  \frac{1}{\pi} \Im m \ \frac{a_{G}}{G(x+i 0^+ )} = \ \frac{a_G \ \rho_{G}(x)}{( \pi \rho_{G}(x))^2 + \left(  \Re e \ G(x)\right)^2 }. \label{inverse}
\eeq
The real part can be found either by taking the Hilbert transform, 
\beq
\Re e \ {\Sigma}(x)= {\cal H}[\rho_\Sigma](x), \label{inverse2}
\eeq
or more directly as
\beq
\Re e \ {\Sigma}(x) =  x- \hat{E} - \Re e \ \frac{a_{G}}{G(x+i 0^+ )} =   x- \hat{E} -  \ \frac{a_G \ \Re e \  G(x)}{( \pi \rho_{G}(x))^2 + \left(  \Re e \ G(x)\right)^2 }.
\eeq

\subsection{\bf Spectral representation for the  ECFL self energies:}
 For the ECFL  Greens function in \disp{twog2},  we set $a_G = \left( 1- \frac{n}{2} \right)$ and write $\hat{E} \to \xi$ representing the single particle energy measured from the chemical potential. We start with the expression:
 \beq
G(\xi,z) = \frac{1}{z-\xi - \Phi(z)} \times \left\{ \left( 1- \frac{n}{2} \right) + \Psi(z) \right\}, \label{eq-ecfl-again}
\eeq 
and express it in terms of the two spectral functions $\rho_\Psi$ and $\rho_\Phi$ \cite{fn3}. 
  We can write  spectral function $\rho_{G}$:     
\beq
\rho_{G}(\xi, x) =   \ \frac{\rho_{\Phi}(x)}{( \pi \rho_{\Phi}(x))^2 + \left(x- \xi-  {\cal H}[\rho_\Phi](x)\right)^2 } \times  \left(   \left( 1- \frac{n}{2} \right) +
 \frac{\xi - x}{\Delta(\xi,x)}+\eta(\xi,x)  \right), \label{spectral-2}
\eeq
where    $\Delta(\xi,x) $ and the term $\eta$  are defined as:
\barray
\Delta(\xi,x) &=& -\frac{\rho_{\Phi}(\xi,x)}{\rho_\Psi(\xi,x)},  \label{delta} \\
\eta(\xi,x) & =&   {\cal H}[\rho_\Psi](\xi,x) +  \frac{1}{\Delta(\xi,x)}  {\cal H}[\rho_\Phi](\xi,x). ~
\label{eta}
\earray
The real part of $G$ is also easily found as
\beq
\Re e \ {G}(\xi, x) =   \ \frac{( \left( 1- \frac{n}{2} \right) + {\cal H}[\rho_\Psi](\xi,x))\left(x- \xi-  {\cal H}[\rho_\Phi](\xi,x)\right) - \pi^2 \rho_{\Psi}(\xi,x)\rho_{\Phi}(\xi,x) }{( \pi \rho_{\Phi}(\xi,x))^2 + \left(x- \xi-  {\cal H}[\rho_\Phi](\xi,x)\right)^2 }. \label{spectral-6}
\eeq
 Thus given the ECFL form of the Greens function, we can calculate the Dyson Schwinger form of self energy in a straightforward way using the inversion formula \disp{inverse} and \disp{inverse2}. The inverse problem of finding $\Phi$ and $\Psi$ from a given $\Sigma$ or $G$ is expected to be ill defined, as discussed above for finite systems.

The   first Fermi liquid factor in \disp{spectral-2} has a peak at  the Fermi liquid  quasiparticle frequency $\E_k$ for a given $\xi_k$ given as   the root of
  \beq
 \E_k - \xi_k-  {\cal H}[\rho_\Phi](\xi_k ,\E_k )=0,  \label{quasiparticle}
  \eeq
however $\rho_G$ itself has a slight shift in the peak due to the linear $x$ dependence in the numerator. This is analysed in detail in the next section  for a model self energy.   
At this solution $x(\xi)$, \disp{spectral-6} gives  a relation:
 \beq
\rho_{\Psi}(\xi_k, \E_k) = - \rho_{\Phi}(\xi_k, \E_k) \times \Re e \ G(\xi_k, \E_k). \label{ecfl-reals1}
\eeq
 
\subsection{High Dimensional ECFL  Model with  $\vec{k}$ independent self energies and its Dyson representation:}
 
  In this section we illustrate the two self energies and their relationships in the context of the recent work on the ECFL \refdisp{ECFL}, and in \refdisp{gweon-ecfl}. Here  we study  a model Greens function,  proposed in \refdisp{ECFL}   for  the \tJ model, that should be suitable in high enough dimensions. It is sufficiently simple so that most calculations can be done analytically. The model Greens function satisfies the Luttinger Ward sum rule\cite{luttinger} and thereby maintains  the Fermi surface of the Fermi gas,  but   yields spectral functions that are qualitatively different from the Fermi liquid. This  dichotomy is possible  since it corresponds to a simple approximation within a formalism that is very  far from the standard Dyson theory, as explained in the previous sections. Our aim in this section is to take this model Greens function of the ECFL and {\em to express it in terms of the Dyson self energy} so as to provide a greater feel for the model.
 
\begin{figure}[t]
\includegraphics[width=5in]{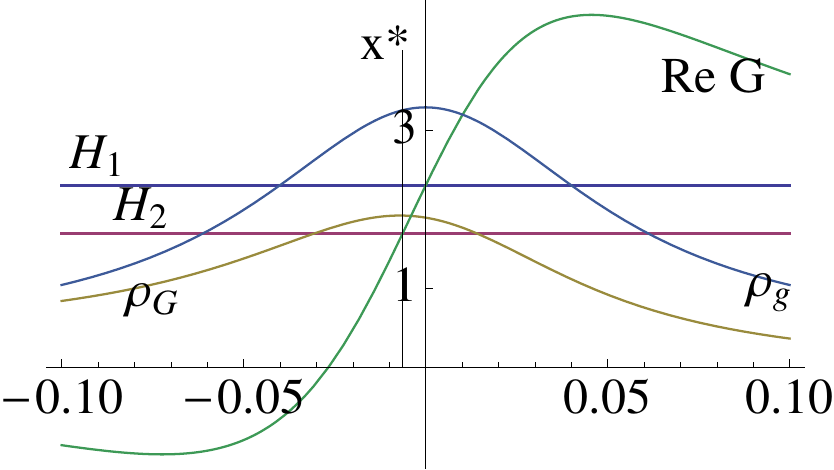}
\caption{  
The density $n=.85$, temperature $T=600$K, $\Delta_0=0.0786$ and  parameters are Set(I) of \disp{datasets}. At this rather high temperature, we can see the details of the spectral shape clearly. The vertical line is at $x^*=E^*_{k_F}$, this energy  is the location of the peak of the physical spectral function $\rho_G$ as marked. Its leftward (i.e. red) shift relative to the Fermi liquid peak at the chemical potential is clearly seen.
The two horizontal lines specify the magnitude of the $\Re e \ G(0,x)$ at $x=0$ (H1) and $x= E^*_{k_F}$ (H2).  The line H1 is at  height $n^2/(4 \Delta_0) $  and   H2 is at height   $n^2/(4 \Delta_0)(1-Z_{k}/2)$. 
  }
\label{Fig_a}
\end{figure}

 Here the two self energies are taken to be  frequency dependent {\em but momentum independent}, and using the formalism of \refdisp{ECFL}  become  related through $\Delta_0$,  an important  physical parameter of the theory:
 \beq
 \Psi(z)= -  \frac{  n^2}{4 \Delta_0}  \Phi(z). \label{delta-def}
 \eeq 
{\col  The physical meaning of  $\Delta_0$ as the mean inelasticity of the auxiliary Fermi liquid (aux-FL)  is emphasized in \refdisp{ECFL}, and follows from \disp {delta-eq}. } 
 Thus $\rho_{\Psi} =  -  \frac{  n^2}{4 \Delta_0} \rho_{\Phi}$, and hence we get the simple result \cite{fn1} :
 \beq
 G(\xi_k, z)= g(\xi_k,z ) \left[  \left\{ 1- \frac{n}{2}  \right\} -  \frac{  n^2}{4 \Delta_0} \Phi(z) \right].
 \eeq
The auxiliary Fermi liquid has a Greens function $g^{-1}(\xi_k,z)= z - \xi_k - \Phi(z)$, where $\xi_k$ is the electronic energy at wave vector $k$ measured from the chemical potential $\mu$, and therefore we may write the model Greens function as 
\beq
G(\xi_k, z  )= \frac{  n^2}{4 \Delta_0} +  \left( \frac{  n^2}{4 \Delta_0} \right) \ \frac{\xxi+\xi_k-z}{z - \xi_k - \Phi(z)} 
\eeq
where
\beq
\xxi=  \Delta_0 \ \frac{4}{n^2} \  ( 1- \frac{n}{2}) . \label{def-delta0}
\eeq
With $\Gamma(x)= \pi  \rho_{\Phi}(x)$, $\Re e\ \Phi(x+i 0^+)= {\cal H}[\rho_\Phi](x)$ and  $\varepsilon(\xi_k,x) \equiv  \left( x -\xi_k -   {\cal H}[\rho_\Phi](x) \right)$, we can express the spectral function and the real part of the Greens function as
\barray
\rho_G(\xi_k,x)& =&   \left(\frac{n^2}{ 4 \pi \Delta_0} \right)\  \frac{\Gamma(x)}{\Gamma^2(x)+ \varepsilon^2(\xi_k,x)}\left\{ \xxi + \xi_k -x \right\}, \label{spectral-8} \\
\Re e \ {G}(\xi_k, x) & =&  \left(\frac{n^2}{ 4  \Delta_0} \right)\ \left[ 1+   \frac{  \varepsilon(\xi_k,x) (\xxi+\xi_k-x) }{\Gamma^2(x)+ \varepsilon^2(\xi_k,x)} \right] .
 \label{spectral-7}
\earray
The linear frequency term in braces in \disp{spectral-8}  is termed the caparison factor in \refdisp{ECFL} and leads to significant features of the spectrum as discussed below.
For completeness we note 
the  auxiliary Fermi liquid part of the problem as
\beq \rho_{\GH}(\xi_k,x)  = \frac{1}{\pi} \frac{\Gamma(x)}{\Gamma^2(x)+ \varepsilon^2(\xi_k,x)}. \label{auxspectra}
\eeq
In \figdisp{Fig_a} we plot the above  three functions for a model system described more fully in Section( \ref{numex}).   
\subsection{EDC or constant wave vector scans  and energy  dispersion.}
{\col We first study the peak structure corresponding to fixing  $\vec{k}$ the wave vector and hence $\xi_k$, and sweeping the energy $x$. These give rise to  the energy distribution curves, i.e. the  EDC's.} 
The aux-FL part has a peak at  $x=\E_k$ for a given $\xi_k$, as in standard FL theory from solving for the roots of  \disp{quasiparticle}. For $k \sim k_F$ we find   $\E_k=\xi_k Z_{k}$ with  the momentum independent self energy $\Phi(z)$,
where
\beq
Z_{k}=\lim_{x\to \E_k} (1- \partial \ \Re e \Phi(x) /\partial x)^{-1}.
\eeq
Expanding around this solution we write 
\beq
\varepsilon(\xi_k,x)  \sim \frac{1}{Z_{k}} (x- Z_{k} \ \xi_k ),
\eeq

We will also write $\Gamma_k\equiv\Gamma(x)/_{x \to \E_k}$ at the FL quasiparticle location, where we expect for the Fermi liquid $\Gamma_k \sim c_1(k-k_F)^2+ c_2 T^2$, with suitable values as described more fully in Section( \ref{numex}).  At this value, we have the identity $\Re e \ {G}(\xi_k, \E_k)  = \frac{n^2}{ 4 \Delta_0}$ as remarked above. As a consequence,   in \figdisp{Fig_a} the intersection of the line H1 and the vertical $y$ axis also coincides with the value of $\Re e \ G$ at the chemical potential.
To elucidate  the line shape  of the ECFL, we  start with the  FL solution and perturb around it  to find the corrected  location of the peaks in the full spectral function.
\beq
\rho_G^{Peak}(\xi_k,x) = \frac{1}{\pi} \frac{Z_{k}^2 \ \Gamma_k}{Z_{k}^2 \ \Gamma_k^2+(x-\E_k)^2} \ \frac{n^2}{ 4 \Delta_0} \ \left\{ \xxi + \xi_k -x \right\}.\label{gpeak-1}
\eeq
\begin{figure}[h]
\includegraphics[width=5in]{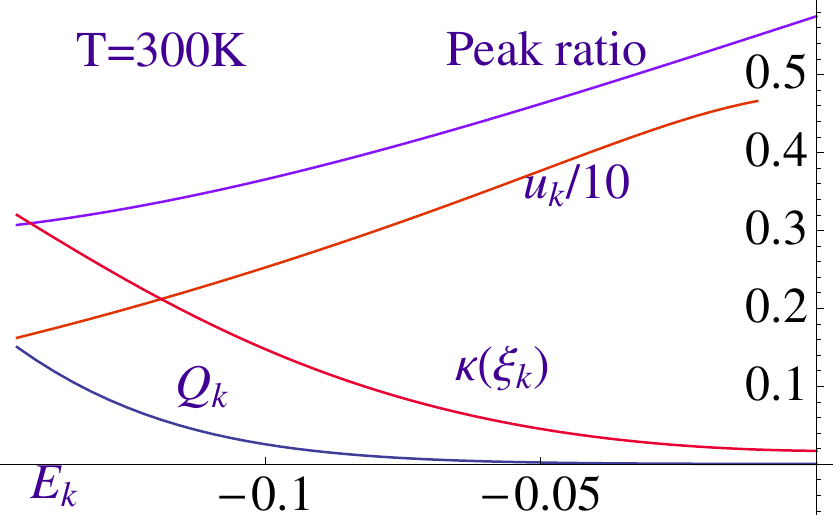}
\caption{  
The density $n=.85$,  $T=300$K, $\Delta_0=0.0678$ and parameters are from Set(I) in \disp{datasets}. 
The various dimensionless variables shown against  $\E_k$  are the  peak ratio from \disp{peak-fl}, the variable $u_k$ (scaled by $10$) from \disp{convenient}, the skew asymmetry factor $\kappa(\xi_k)$ from \disp{skew} and the variable $Q_k$ from \disp{qk}.
}
\label{Fig_q}
\end{figure}
Similarly the real part is found
\beq
\Re e \ G^{Peak}(\xi_k,x) =  \frac{n^2}{ 4 \Delta_0}  \left[1 +  Z_{k} \frac{(x- \E_k) \left\{ \xxi + \xi_k -x \right\} }{
Z_{k}^2 \  \Gamma_k^2+(x-\E_k)^2}  \right]. \label{gpeak-2}
\eeq

We introduce the following convenient positive  variable $u_k$:
\beq
\sinh{u_k} \equiv \frac{\xxi+ \xi_k - \E_k}{  Z_{k} \Gamma_k }, \label{convenient}
\eeq
so that {\col near the Fermi energy} and at low $T$ the small $\Gamma_k \sim T^2 $ drives it  to a  large and positive value {\col i.e. $\exp{u_k} \to 1/T^2$ }.  At higher binding energies, $u_k$ decreases towards zero, as discussed below. We will also
define a dimensionless variable $Q_k$ below in  \disp{qk} that depends on $u_k$ only and determines the shape of the peak. 
To analyze the shape at a given $\xi_k$,  we introduce {\col a dimensionless} energy variable $\epsilon$ through the relation
\beq
x= \E_k + Z_{k} \Gamma_k \ \epsilon,
\eeq
where we must require that $|\epsilon| \sim 1$ for the expansion around the FL peak to be valid.
The spectral function is expressible as
\beq
\rho_G^{Peak}(\xi_k, \epsilon) = \rho_G^{*}(k) \left[ \frac{\sinh(u_k)- \epsilon}{1+ \epsilon^2}\right] \ 2 e^{-u_k}, \label{peak-2}
\eeq
and
\beq
\Re e \ G^{Peak}(\xi_k,x) =  \frac{n^2}{ 4 \Delta_0}  \left[1+Z_{k} \ \epsilon \  \frac{(\sinh(u_k) -\epsilon)}{1+\epsilon^2}\right].
\eeq
From \disp{peak-2} we see that  at any $k$, the spectral function peaks at $\epsilon^* \equiv -e^{-u_k}$ with the true quasiparticle peak $E^*_k$ corrected from the Fermi liquid value $\E_k(= Z_{k} \ \xi_k)$ as: $ E^*_k  \equiv \E_k - e^{-u_k} Z_{k} \Gamma_k, $ at $\epsilon^*=- e^{-u_k}$. Simplifying, we find the EDC energy dispersion or spectrum
\barray
 E^*_k&=&
 \xi_k+\varepsilon_0 - \sqrt{\left[ \varepsilon_0+(1-Z_{k}) \ \xi_k\right]^2+ Z^2_{k}\Gamma^2_k} \ .
  \label{qpcorr}
\earray
We provide examples of this dispersion later in \figdisp{dispersion}.

For a given $\xi_k$, the magnitude of the spectral function at this peak  is given by
\beq
 \Re e \ G^*(k)=  \frac{n^2}{ 4 \Delta_0}  \left[1-\frac{1}{2} Z_{k} \right],\;\;\;\;\rho_G^{*}(k)= \frac{n^2 \ Z_{k} }{8 \pi  \Delta_0} e^{u_k}. \label{peak-ecfl}
\eeq
The magnitude of $\Re e \ G^*(k)$ is  a little smaller than the value $ \frac{n^2}{4 \Delta_0}$ arising at the FL solution $\epsilon=0$. In \figdisp{Fig_a} this is reflected in the line H2 that lies a little below H1 \cite{fn4}.

The peak value $\rho_G^{*}(k)$ falls off  with $\xi_k\ll 0$, and is always smaller relative to  the peak of the aux-FL peak value $ \rho^*_g(k)$. The ratio of the two peak values is  given by
\beq
\frac{\rho^*_G(k)}{\rho^*_g(k)} 
=\frac{n^2 \ Z_{k} \Gamma_k}{8   \Delta_0} e^{u_k}. \label{peak-fl}
\eeq
We see below numerical examples of these functions. \figdisp{Fig_q} illustrates the peak ratio and other features for a typical set of parameters.

\subsection{MDC or constant energy scans and energy dispersion.}
It is also useful to study the momentum distribution curves ``MDC'' obtained by fixing the energy $x$ and scanning the energy $\xi_k$ \cite{mdc}.
 In the  model of a $\vec{k}$ independent self energy, this is a particularly convenient strategy, and hence maximizing \disp{spectral-8} at a fixed $x$ we find the MDC energy dispersion or spectrum:
\beq
\xi^*(x) = x- \varepsilon_0 + \sqrt{\Gamma^2(x)+ (\varepsilon_0 - \Re e \Phi(x))^2}. \label{mdc}
\eeq
Thus $\xi^*(x)$ is the peak  position of $\xi_k$  in constant energy scans, whereas  $E^*_k$ in \disp{qpcorr} represent peak position of  energy at a fixed $\xi_k$
It is amusing to compare this with \disp{qpcorr}. Unlike \disp{qpcorr}, this formula is valid at all energies, not just near the chemical potential where the two agree closely.  We will see below  in \figdisp{dispersion} that this function is multivalued in a range of values of energy $x$  leading to  characteristic  features of the spectrum.
   
\subsection{ Numerical example of High Dimensional ECFL model: \label{numex}}
In this section we use a  a rectangular band with height $1/(2 W)$ and width $2 W$,    and take  $W=.86$ eV (i.e. $10^4 $K) as a typical value. 
   In \refdisp{gweon-ecfl} a more realistic band structure is used as described in detail there. The   model   for the Fermi liquid introduced in \refdisp{ECFL} (Eq(24)) is given by the expression:
  \beq
\Gamma(x)= \pi \rho_{\Phi}(x) = \pi  C_\Phi \{ x^2+  \tau ^2  \}  e^{- C_\Phi \{ x^2+ \tau^2 \}/\omega_c } +\eta,  \label{model-sigma} 
\eeq
with $\tau= \pi k_B T$. We have added a scattering width $\eta$ as in \refdisp{gweon-ecfl}, in order to account for scattering by off planar impurities.
  The  real part of the self energy is found from the Hilbert transform of $\rho_\Phi(x)$, and  is given by:
\beq  
\Re e \Phi(x) =     C_\Phi \pi  (x^2+ \tau ^2) e^{- C_\Phi \{ x^2+ \tau ^2  \}/\omega_c}  \  \mbox{Erfi}(x /  \sqrt{\omega_c})- C_\Phi  \ x \ \sqrt{\pi \omega_c}  \  e^{- C_\Phi \tau^2  /\omega_c } ,
\eeq
where $  \mbox{Erfi}(x)= \frac{2}{\sqrt{\pi}} \int_0^x e^{t^2} \ dt$ is the imaginary error function. 
A numerically small correction arising from $\eta$ is dropped for brevity.
{\col
\subsection{Typical parameters} 
The same model is also used in the fit to experiments in \refdisp{gweon-ecfl} with a slight change of notation given by writing $C_\Phi \to \frac{1}{\pi \Omega_0}$ and 
$\omega_c\to \frac{\omega_0^2}{\pi \Omega_0}$, in terms of the high and low frequency cutoff  frequencies $\omega_0$ and $\Omega_0$. We use two sets of standard parameters
\barray
\mbox{Set-I:}&\;\; C_\Phi=1 \ eV^{-1}, \;\; \omega_c=0.25 eV\;\; \mbox{or}\;\; \omega_0= 0.5 eV,\;\;\Omega_0= .318 eV \nn \\
\mbox{Set-II:}& \;\; C_\Phi=2.274 \ eV^{-1}, \;\; \omega_c=0.568 eV\;\; \mbox{or}\;\; \omega_0= 0.5 eV,\;\;\Omega_0= .14 eV.  \label{datasets}
\earray
 Set-I was used in \refdisp{ECFL} for schematic plots employing a simple band density of states $g_B(\varepsilon)= \frac{1}{2 W} \Theta(W^2-\varepsilon^2)$.
Set-II was used in \refdisp{gweon-ecfl} employing a more elaborate dispersion described therein,  to successfully fit data on various  high temperature superconductors at optimal doping.  The value of $\eta$ is displayed in different plots.
}
  In \disp{auxspectra} the spectral function $\rho_g$ of the aux- FL is defined.  The chemical potential is fixed  by the number sum rule  with $\xi = \epsilon - \mu$
  \beq\frac{n}{2}=  \int_{-\infty}^{\infty} dx \   f(x)  \int d \epsilon \ g_B(\epsilon)  \  \rho_{\GH}(\epsilon- \mu,x), \label{number-sumrule}
  \eeq
  where $f(x)=(1+ e^{\beta x})^{-1}$ is the Fermi function.  We now  write the contributions from extreme correlations that are described in \refdisp{ECFL}. The inelastic  energy scale $\Delta_0$  is found from the sum rule:
\beq
\Delta_0=    \int_{-\infty}^{\infty} dx \   f(x)  \int d \epsilon \ g_B(\epsilon)  \  \rho_{\GH}(\epsilon- \mu,x)  \ \{ \epsilon - \mu - x   \}. \label{delta-eq}
\eeq
Thus at a given density and temperature $n,T$, the model has only two  parameters $\omega_c$ and $C_\Phi$ so that $\Delta_0$ is  fixed from \disp{delta-eq}. 
  We study the details of the spectra next.

{\col 
\section{ The spectral characteristics of the  High Dimensional  ECFL  model \disp{spectral-8} }
\subsection{Global view of the spectral function}

 \begin{figure}[h]
\includegraphics[width=3.6in]{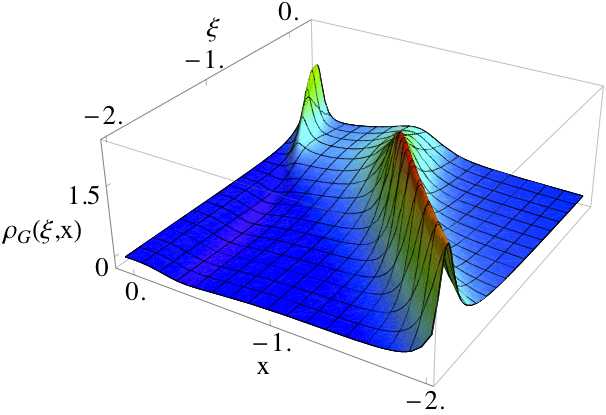}
\includegraphics[width=2.8in]{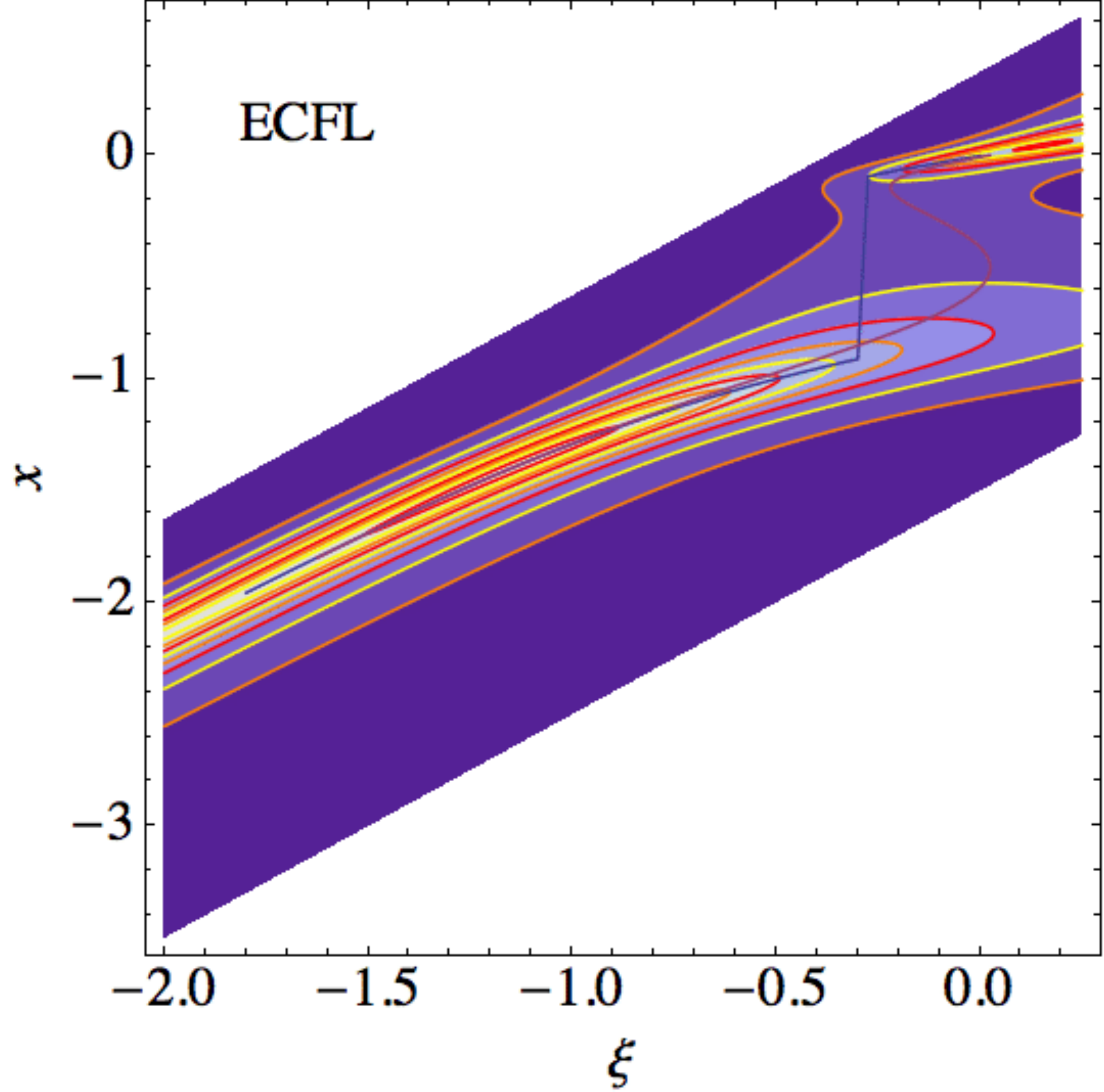}\\
\includegraphics[width=3.6in]{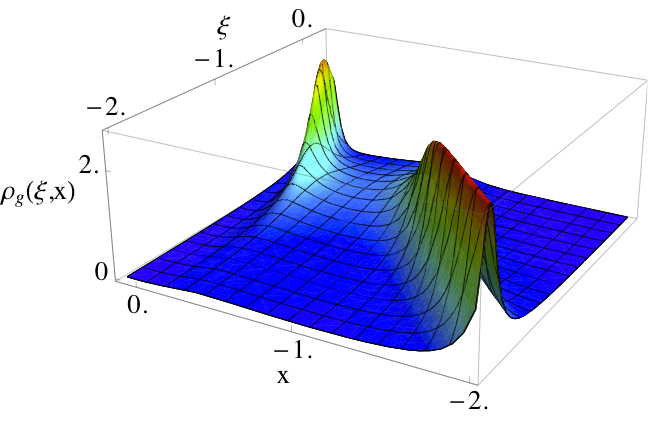}
\includegraphics[width=2.8in]{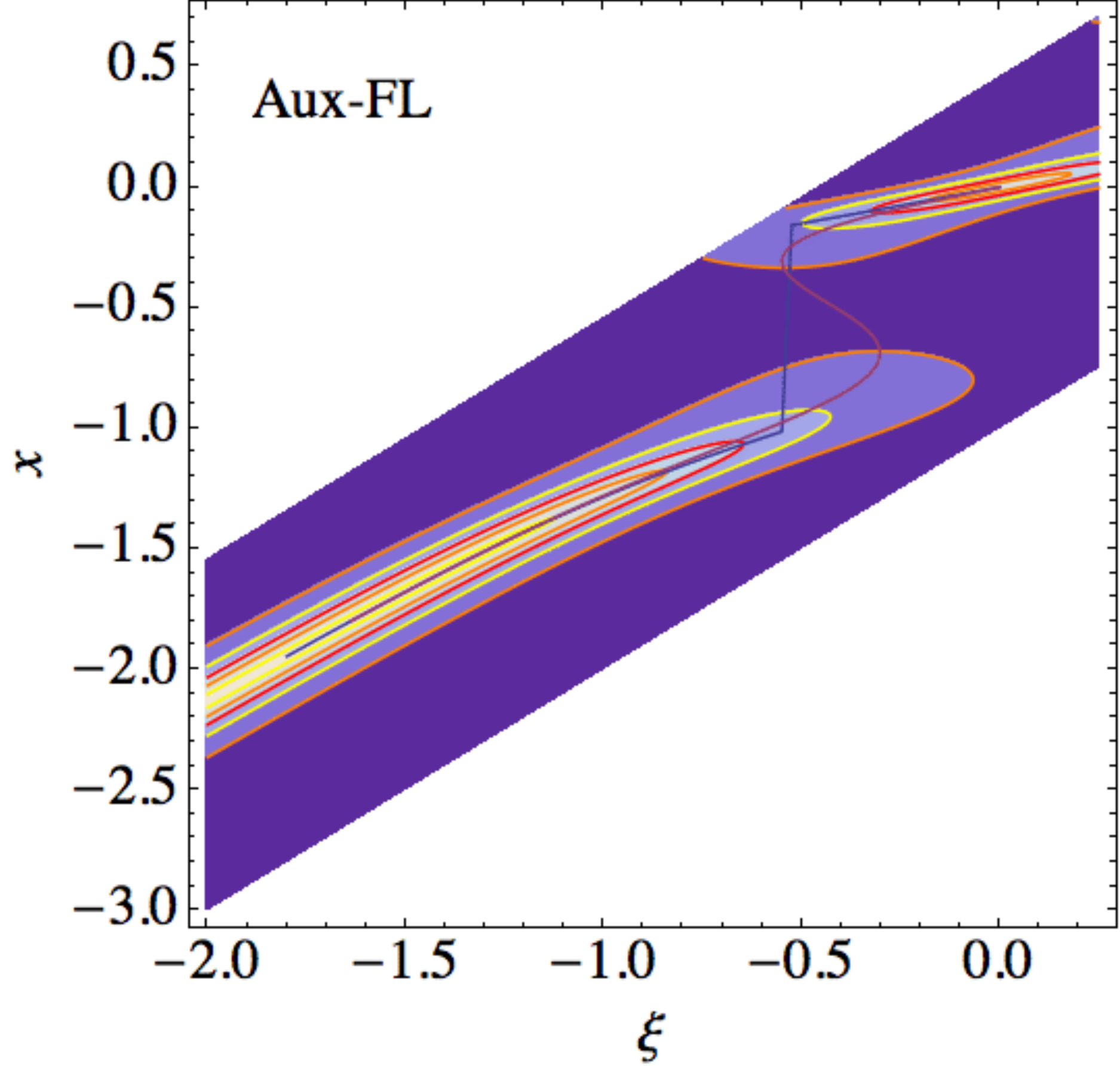}
\caption{  
The ECFL  (top left)  and the auxiliary FL spectral function (bottom left) at density $n=.85$,  $T=180$K,  $\Delta_0=0.12$ eV, $\eta =0.12$ eV, and the other parameters are from Set(II) in \disp{datasets}. Here $\xi$ and $x$ are in units of eV.  In the ECFL curve on left, it is seen that  the excitations near the Fermi energy become broad and dissolve into the continuum  at an energy $\sim -.2 $ eV, and reappear as sharp modes at a deeper binding energy. In the auxiliary FL,  the excitations near the Fermi energy  remain sharp and extend to lower energies than in the ECFL curves. The contour plots  of the same functions in  the right panel (top ECFL and bottom auxiliary FL)   gives a complementary perspective of the spectrum. The two superimposed solid lines at top right are from  Curves (I) and (II) of \figdisp{dispersion} and at  bottom right  Curves (III) and (IV) of \figdisp{dispersion}.  }
\label{kink}
\end{figure}
We  display in Fig.~(\ref{kink}) the spectral function for  the ECFL model \disp{spectral-8} in 3-d plots  and contour plots. Two distinct perspectives of the spectrum are found in the figure from the 3-dimensional plot and the contour plot. In both of these we see that the excitation are sharply defined only for a certain range near the Fermi energy, and then merge into the continuum. At higher binding energies, the spectrum again looks quite sharp. For comparison, in Fig.~(\ref{kink}) we also display the aux-FL spectral function.  We note that the aux FL spectra also become sharp at higher binding energies. This sharpening is modeled by the Gaussian in \disp{auxspectra},  its basic origin  is  the decrease in the weight of physical processes capable of  quasiparticle damping as we move towards the band bottom. In order to look more closely at the low energy part of the spectrum of the aux-Fl and the ECFL, we show in \figdisp{comparecontour} the contour plots of both next over a smaller energy range.
\begin{figure}[h]
\includegraphics[width=3.2in]{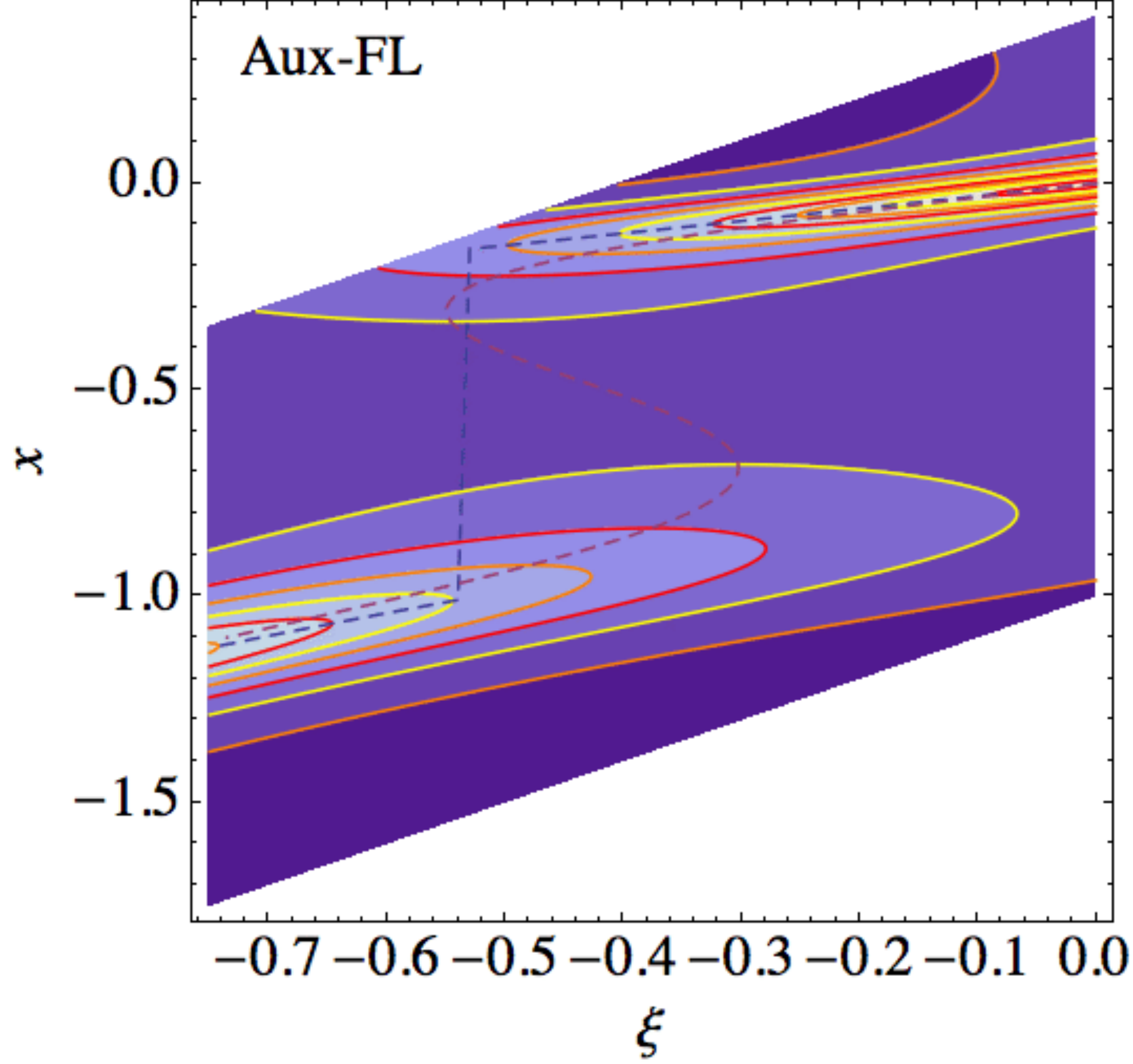}
\includegraphics[width=3.2in]{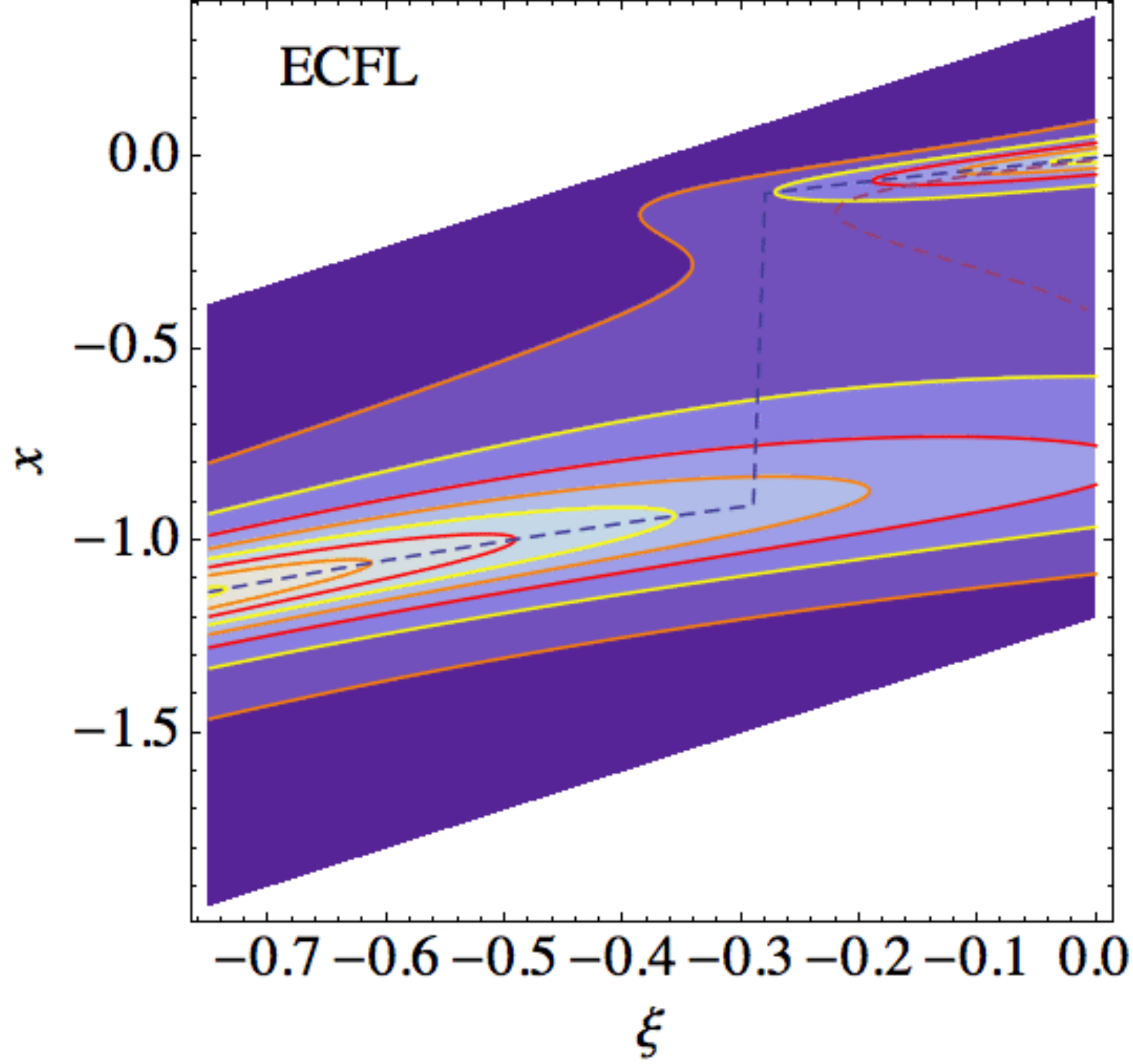}
\caption{  
The contour plots of  the aux FL (left) and the ECFL model from \figdisp{kink} (right) with the same parameters as in \figdisp{kink} but over a smaller energy window.  
We superimpose the constant wave vector  dispersion and MDC dispersion, with a value of $\eta=.12$  common to the contour plots. The energy scale of the feature near the chemical potential is considerably reduced in the ECFL, and the ``jump" in the EDC  dispersion occurs at roughly half the corresponding energy in the aux-FL.
 }
\label{comparecontour}
\end{figure}
We see that viewed in this rather broad sense, dispersions of   the aux FL and the model ECFl spectra share many characteristics,  with somewhat different energy scales.  However  there are crucial differences that emerge when we look at the distribution of spectral weight that arises in the ECFL, where the caparison factor in \disp{spectral-8} pushes weight to higher binding energies. This is reflected most significantly in the line shapes  that we study below. } Since we use the momentum independent self energy for the aux-FL in this model calculation, we obtain very detailed EDC and MDC plots below. However, it must be borne in mind that refined calculations within the ECFL framework must necessarily introduce some momentum dependence, and hence several details are likely to change, in particular the structure far from the chemical potential would change somewhat more. Our view is that  this caveat apart, it is very useful to take the \disp{spectral-8} seriously since it gives a simple framework to  correlate different data.

\subsection{Dispersion relations in EDC and MDC}
\begin{figure}[h]
\includegraphics[width=5in]{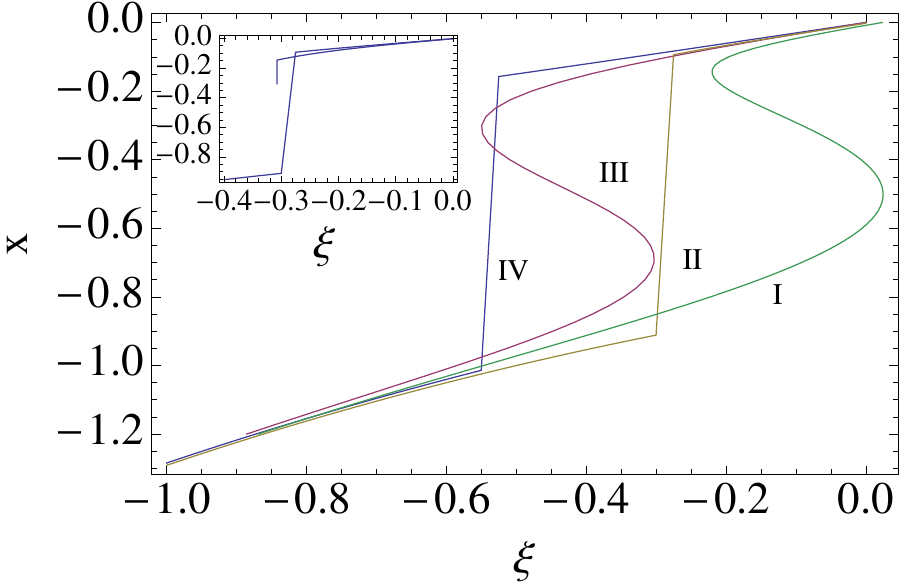}
\caption{ Energy dispersion curves in the ECFL and the aux FL models. Here the parameters are from Set(II) in \disp{datasets}, with $n=0.85$, $T=180$K. With $\eta=0.12$,  curves I and II are the peaks in constant wave vector and constant energy  scans of the spectral function \disp{spectral-8}, and curves III, IV are corresponding figures for the aux FL in \disp{auxspectra}.  The inset  compares \disp{qpcorr}  (the truncated curve) with the exact locus found by numerical maximization.
}
\label{dispersion}
\end{figure}

In  \figdisp{dispersion} the  EDC dispersion relation  (i.e. locus of peaks  of  the spectral function at fixed $\xi$, found by numerical maximization),  is plotted versus $\xi$ along with the MDC spectrum  \disp{mdc}. We recall that the latter expression  is exact at all $\xi, x$, whereas  \disp{qpcorr}  is not quite exact for the EDC dispersion.  For comparison we also show the corresponding figures for the aux- FL  spectral function in \disp{auxspectra}, with the same parameters.
 The dispersion relations \disp{qpcorr}  is displayed in the inset of  \figdisp{dispersion}, where it is  compared  with the result of numerically maximizing the spectral function at a fixed $\xi$.  We see that \disp{qpcorr} is only  good for a range of energies near the Fermi energy.
We see that both sets of spectra for the aux-FL as well as the  ECFL model exhibit similar global features, but with different scales of energy. In both cases,  the constant energy scans show a jump discontinuity, 
whereas the MDC spectra show a ``S" type or re-entrant type behaviour.
The  origin of the latter is easy to see  in the aux-FL, here    a peak in $ - \Re e \Phi(x)$ occurs at an energy  approximately $2 \omega_0$, so that as $x$ decreases from zero,  $\xi^*_{aux-FL}=x-\Re e \Phi(x)$ goes back up for a certain range. In case of the ECFL, \disp{mdc} shows that for the energy scale $\varepsilon_0$ enters the expression when $\Gamma(x)$ becomes comparable to $\varepsilon_0 - \Re e \Phi(x)$, and the net result is that the reentrant behaviour is pushed to lower binding  energies.

\subsection{The  energy shift  }

The dispersion \disp{qpcorr} corresponds to the ridge near the Fermi energy in Fig.~(\ref{kink}).  At low temperature, since $\Gamma_{k_F} \sim O(T^2) $, the corrected   quasiparticle  energy is always less than $\E_{k}$, so that there is always a leftward (i.e. red) shift of the dispersion, {\col or from the hole (binding) energy point  of view, we may say there is a blue shift.  The peak shift  is given by
\beq
\Delta E_k = E_k^*-  \E_k = (1- Z_{k}) \xi_k+\varepsilon_0 - \sqrt{\left[ \varepsilon_0+(1-Z_{k}) \ \xi_k\right]^2+ Z^2_{k}\Gamma^2_k} ,  \label{e-shift}
\eeq
which is a function of both $T$ and $k$. Close to the Fermi energy this can be written as
\beq
\Delta E_k =  -   \frac{Z^2_{k_F}  \Gamma^2_k}{ 2 \varepsilon_0}. \label{locpeak-2}
\eeq
At the Fermi momentum, this small shift is  seen in \figdisp{Fig_a} where the vertical line through $x^*$ is displaced to the left from the $y$ axis.   As long as  $k\sim k_F$ this shift  is very small $\Delta E_k\sim O(T^4)$, but as $k$ moves away from $k_F$ the shift \disp{e-shift} grows with $\xi_k$.  This departure  makes the   dispersion in \disp{qpcorr} depart significantly from the bare dispersion $\xi_k$ as we move away from $k_F$.  We see from \figdisp{dispersion} that the departure of the EDC peaks from the Fermi liquid are somewhat less pronounced than those of the MDC's, the latter is  operationally called as the low energy kink.  Our calculations therefore predicts   the magnitude of the shift \disp{e-shift} in terms of the energy scale $\varepsilon_0$ and the Fermi liquid parameter $\Gamma_k$.  This energy shift is therefore also a useful method for extracting the fundamental parameter $\Delta_0$ on using \disp{def-delta0}.
}

\subsection{Constant energy cuts or MDC line shapes}
\begin{figure}
\includegraphics[width=2.in]{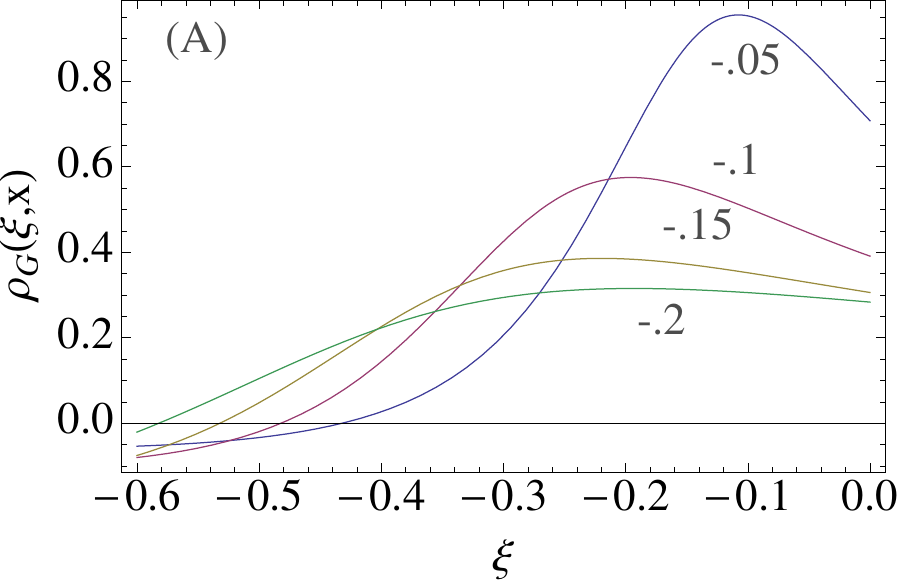}
\includegraphics[width=2.in]{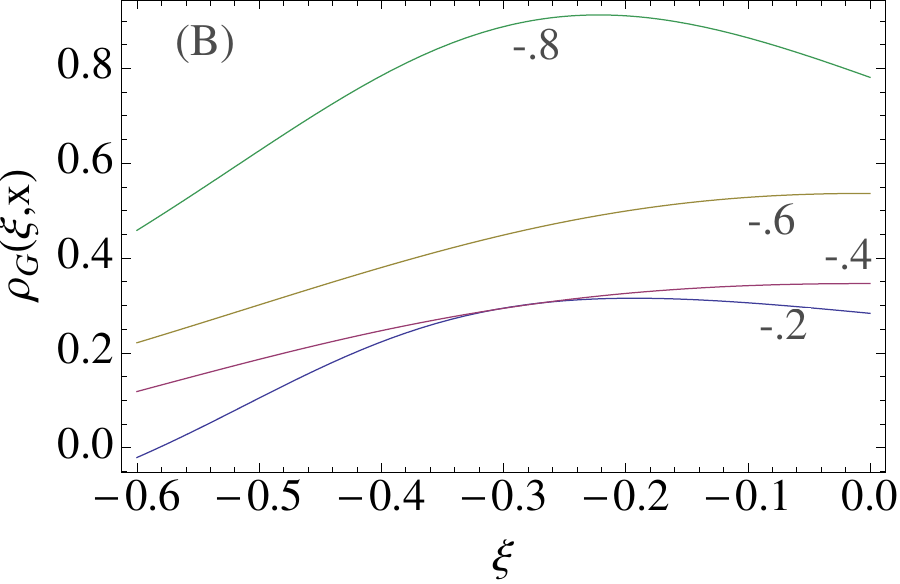}
\includegraphics[width=2in]{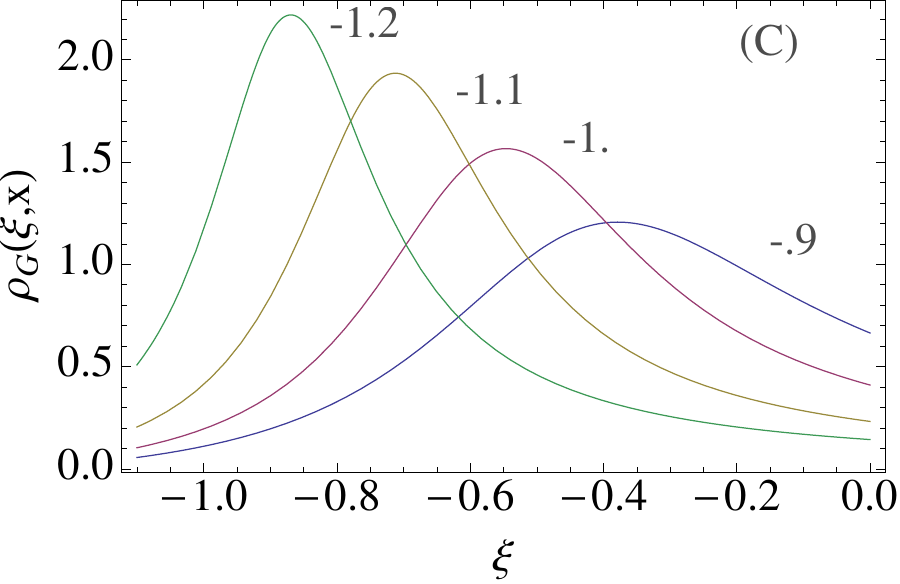}
\caption{ MDC line shapes at different values of  energy $x$ displayed in each curve. Here the parameters are from Set(II) in \disp{datasets}, with $n=0.85$, $T=180$K and $\eta=0.12$. Panel (A) corresponds to $x$ close to the chemical potential. It is  interesting to note   that the curves are skewed to the right, thus mirror imaging   the leftward  skew seen in the constant $\xi$ (EDC) scans below Fig. ~(\ref{Fig_EDC_Spectra} A),  in a comparable range of energy and wave vector. Panel (B) corresponds to the mid energy range, within the re-entrant range of $x$ from \figdisp{dispersion} or  \figdisp{kink}, with the counterintuitive movement of the shallow peak to the right  with increasing $x$. Panel (C) corresponds to the second set of maxima in  \figdisp{kink} far from the  chemical potential, where the curves are quite symmetric. 
  }
\label{Fig_MDC_Spectra}
\end{figure}

We display the MDC line shapes in \figdisp{Fig_MDC_Spectra}. Panel (A) shows the effect of the caparison factor $\left(1- \frac{n}{2} + \frac{n^2}{4 \Delta_0}(\xi -x) \right)$, whereby the curves are {\em  skewed  to the right}, in contrast to the EDC curves that are {\em skewed to the left}. The latter important feature is also seen below in Fig. ~(\ref{Fig_EDC_Spectra} A) and noted in \refdisp{ECFL}. Panel (B) shows the shallow peaks in the ``S like" region of the energy dispersion seen in \figdisp{dispersion}, and panel (C) shows the deep interior region where the peaks are more symmetric.

\subsection{Constant wave vector cuts or EDC line shapes}
\begin{figure}
\includegraphics[width=3.2in]{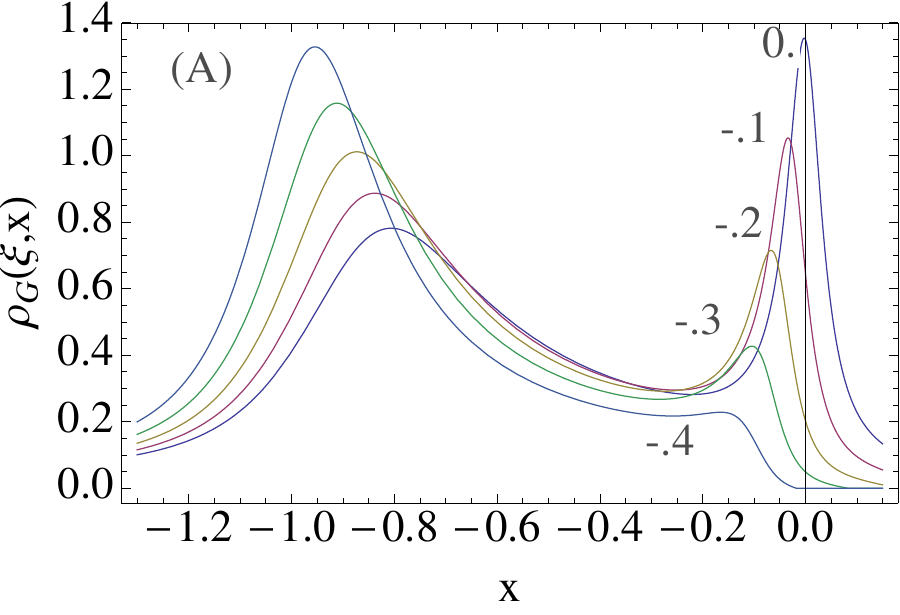}
\includegraphics[width=3.2in]{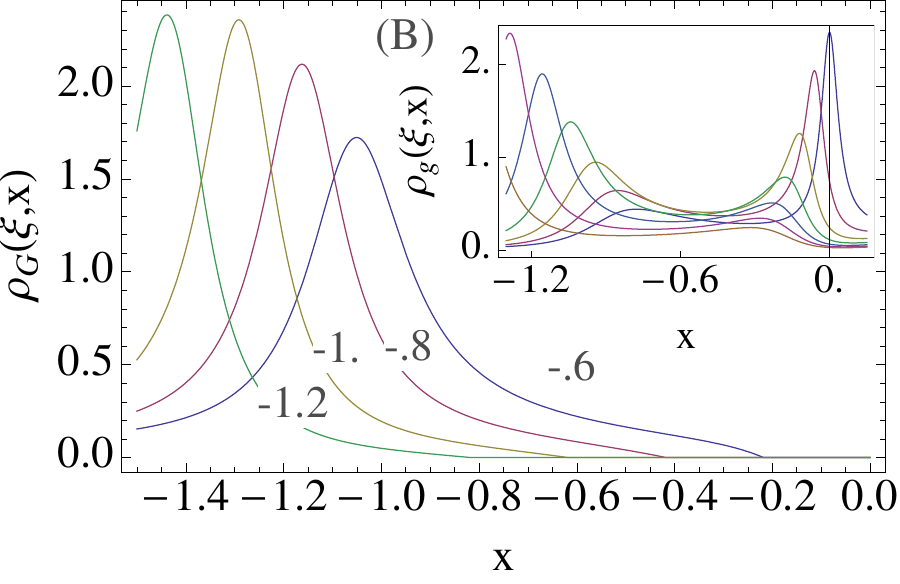}
\caption{ EDC line shapes at different values of  energy $\xi$ displayed in each curve. Here the parameters are from Set(II) in \disp{datasets}, with $n=0.85$, $T=180$K and $\eta=0.12$. Panel (A) corresponds to $\xi$ close to the chemical potential.  Note that the curves are skewed to the left, i.e. a mirror image of the rightward  skew seen in the constant $x$ MDC scans above \figdisp{Fig_MDC_Spectra},  in a comparable range of energy and wave vector. Panel (B) corresponds to the higher energy range, and we see that only one broad  maximum is found at a given $\xi$. The inset of (B) shows the aux-FL constant $\xi$ scans for the same range, here each $\xi$ results in a pair of maxima, originating from the functional form of the self energy in \disp{spectral-8}. 
  }
\label{Fig_EDC_Spectra}
\end{figure}

The spectral function and the real part of the  Greens function are calculated from
\disp{spectral-7} and \disp{spectral-8}.
We display the EDC line shapes in \figdisp{Fig_EDC_Spectra}. Panel (A) gives an overview
of the spectral shapes for wave vectors near the Fermi surface, displaying a left skewed peak that falls rapidly in intensity as it broadens. This behaviour is of great interest since it captures the experimental features in high Tc systems, as elaborated in \refdisp{gweon-ecfl}. Panel (B) shows the spectra at higher binding energies, where a feature at lower energies begins to disperse significantly with $\xi$. It is evident that these two sets of dispersing features correspond to the two branches that are seen in the 3-d plots and contour plots \figdisp{kink}. The inset in (B) shows the behaviour of the aux- FL, where the two features are again seen but with different rates of intensity change.

\begin{figure}
\includegraphics[width=3.in]{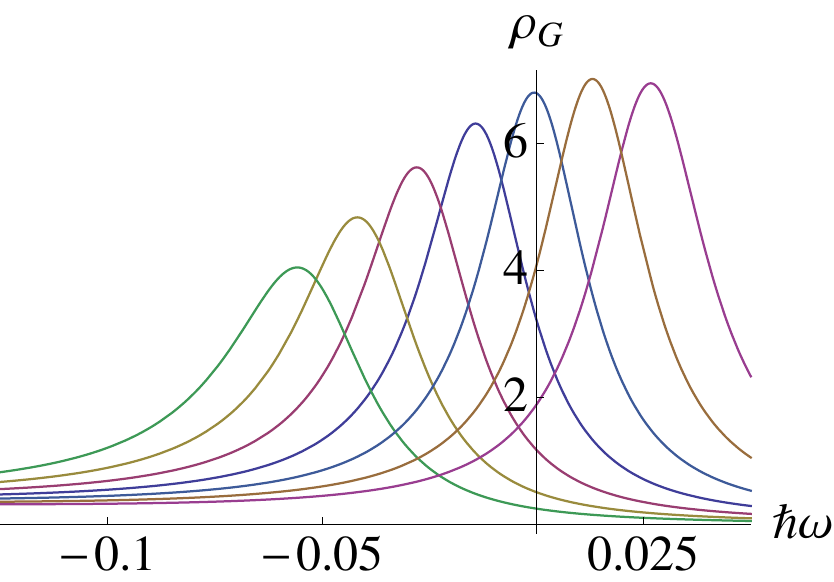}
\includegraphics[width=3.in]{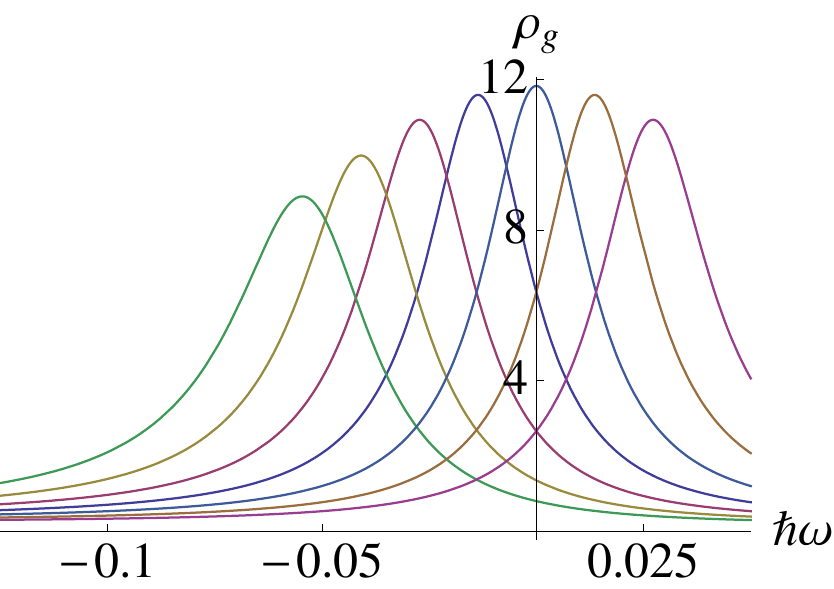}\\
\includegraphics[width=3in]{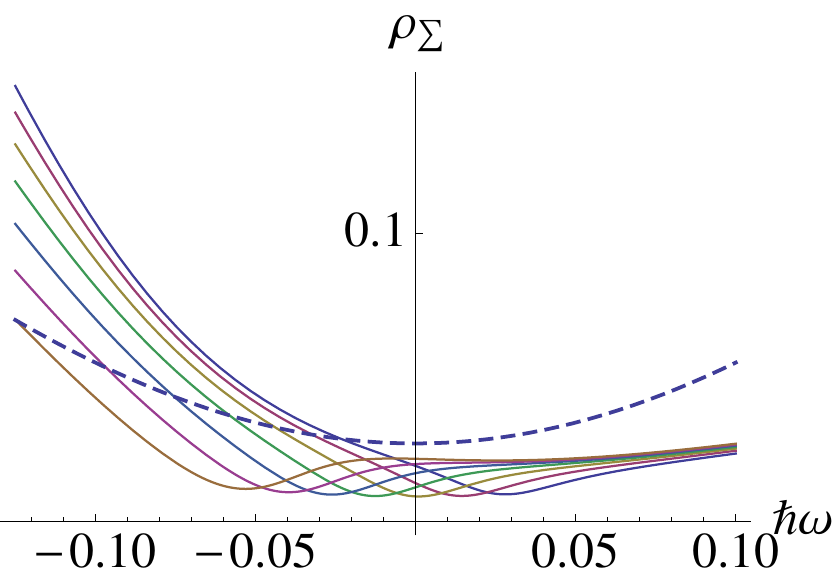}
\includegraphics[width=3in]{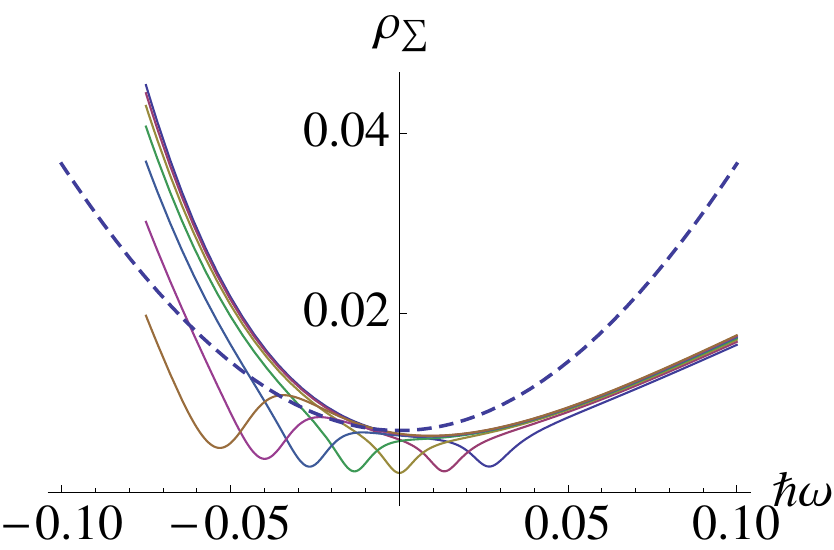}
\caption{{\bf Top Left panel} The density $n=.85$, temperature $T=300$K, $\Delta_0=.0678$ and  other parameters are from Set(I) in \disp{datasets}. From left to right $\rho_{G}(\xi, x)$ for energies in units of eV: $\xi= -0.1, -0.075,-0.05,-0.025,0.,0.025,0.05$. {\bf Top  Right  panel} The spectral function $\rho_g(\xi,x) $  from \disp{auxspectra} corresponding to the same $\xi$ as in the left panel. The difference in the line shapes becomes clear when we  examine the Dyson self energy that produces these curves. {\bf Bottom left panel} The   panel shows the spectral function at $T=300$K for the inferred Dyson  self energy    $\rho_{\Sigma}( x)$ from \disp{inverse} and   \disp{spectral-7}, \disp{spectral-8} for the same energies. The dashed line is the input  Fermi liquid spectral function $\rho_\Phi(\omega)$ at the same temperature. {\bf Bottom Right  Panel} Temperature $T=150$, $\Delta_0=.0642$ and the identical data as in the bottom left panel. }
\label{Fig_b}
\end{figure}

\begin{figure}
\includegraphics[width=5in]{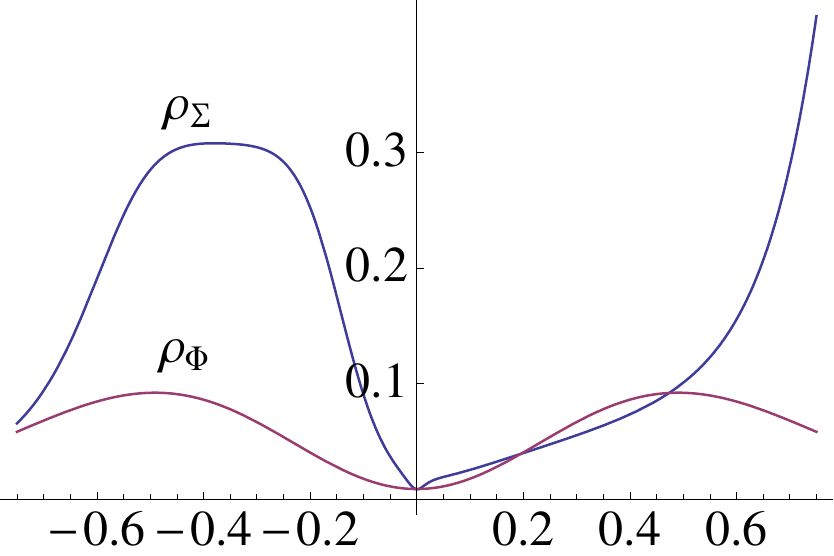}
\caption{ With $n=0.85$ and $T=300K$, and  other parameters are from Set(I) in \disp{datasets}. The  spectral function for the inferred Dyson self energy   $\rho_\Sigma(\xi=0,x)$ using \disp{inverse} and the the input  Fermi liquid spectral function $\rho_\Phi(x)$ over a larger energy range. Note the  distinctive asymmetry  in shape of  $\rho_\Sigma$  below  above the Fermi energy. }
\label{Fig_c}
\end{figure}

We now turn to the task of understanding the reconstructed Dyson self energy that leads to the above electron spectral functions.
In \figdisp{Fig_b} we show the spectral function $\rho_G$ at various values of the energy $\xi$ at $T=300K$. The Fermi liquid spectra at the same values of parameters are also shown for comparison.  The   Dyson self energy $\rho_{\Sigma}(\xi,x)$  necessary to produce these spectral functions is found  using \disp{inverse} and displayed in \figdisp{Fig_b} at two temperatures. The object $\rho_\Sigma(\xi,x)$ has a distinctive minimum for each $\xi$ that shifts to the left along with the  energy $\xi$ that tracks the peaks in the physical spectral function $\rho_G(\xi,x)$ from \disp{inverse}. It also shows  the asymmetry between energies above and below the chemical potential that we noted at $\xi=0$  in \figdisp{Fig_c}.  At the Fermi energy $\rho_\Sigma(\xi,x)$ is displayed in \figdisp{Fig_c} over a large scale.

\subsection{The reduced line shape function }

An interesting  aspect of the  ECFL model Greens function \disp{spectral-8} is the change in shape of the peaks as we leave the Fermi surface,  so that the quasiparticles become hard to define at some point. This change in shape can be formulated neatly in terms of a single dimensionless parameter $Q_k$ that we now define and explore.
We examine  \disp{peak-2} around its peak  by writing
 \beq \epsilon = \epsilon^* + \cosh(u_k) \bar{\epsilon}, \label{ebars}
 \eeq 
 so that $ \rho_G^{Peak}(\xi_k, \bar{\epsilon}) = \rho_G^{*}(k) \gamma(Q_k,\bar{\epsilon})$, with a characteristic  line shape function $\gamma$ given by 
\beq
  \gamma(Q_k,\bar{\epsilon})= \left[ \frac{Q_k( 1- \bar{\epsilon})}{Q_k (1- \bar{\epsilon})+\bar{\epsilon}^2  \ }\right], \label{peak-3}
\eeq
with
\beq
Q_k=2  \frac{ e^{-u_k}}{ \cosh{(u_k)}}. \label{qk}
\eeq
The parameter $Q_k$ goes to zero near the Fermi surface at low $T$ since $u_k \to \infty$, but at higher binding energies increases   $Q_k \to 2$.
\begin{figure}[h]
\includegraphics[width=6in]{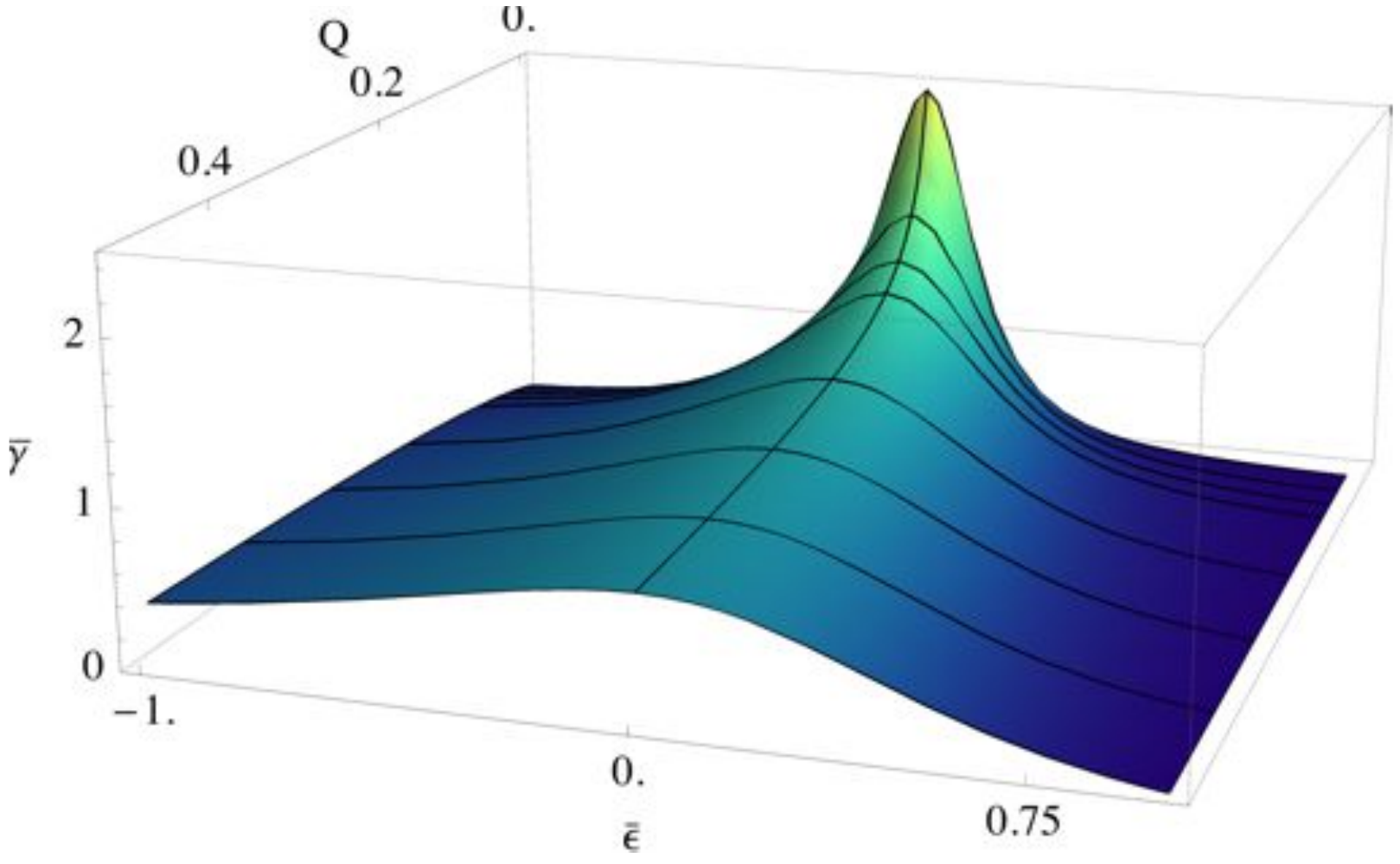}
\caption{  The spectral shapes possible are seen by   plotting the shape function  different values of the parameter $Q$.  In this curve $\overline{\gamma}$ is the $\gamma(Q,\bar{\epsilon})$ of \disp{peak-3}   normalized to  unit area in the  natural interval $[-1,1]$ for the variable $\bar{\epsilon}$.}
\label{Fig_shapes}
\end{figure}

\begin{figure}[h]
\includegraphics[width=5in]{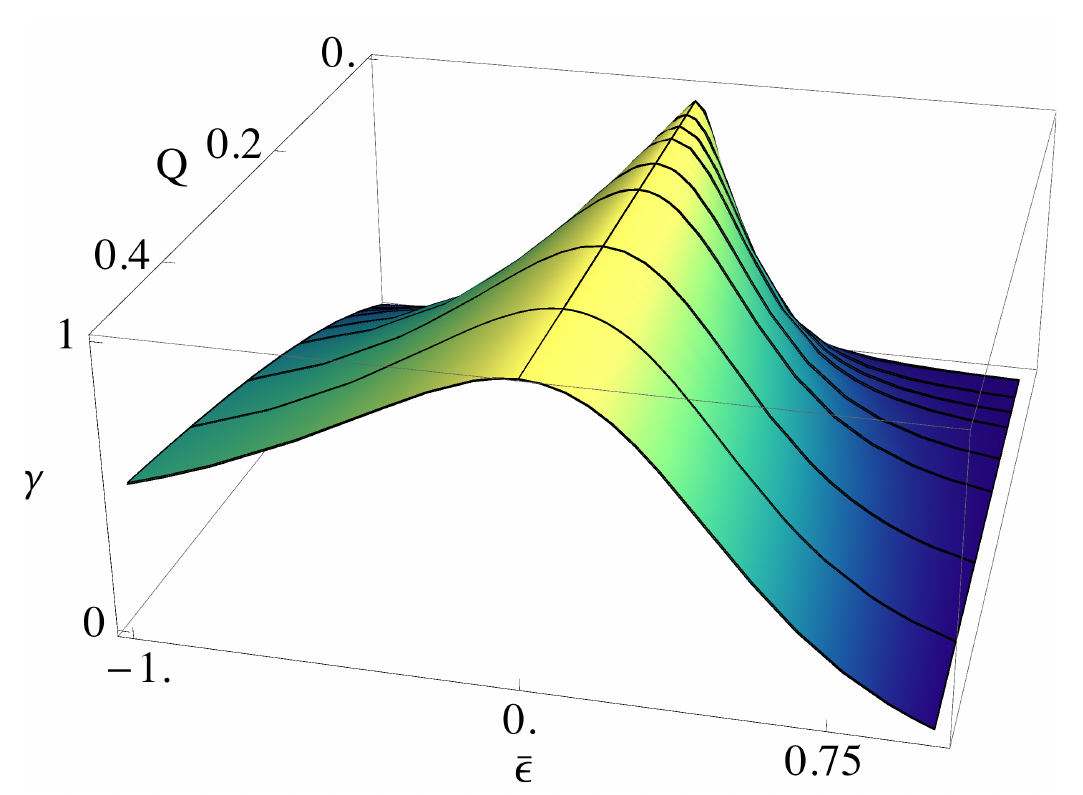}
\caption{  The same shapes as in \figdisp{Fig_shapes} but now $\gamma$ is normalized to unity at the peak as in \disp{peak-3}. The sharp peaks for small $Q\leq .25$ flatten out as $Q$ increases with a left skew asymmetry that is characteristic of this functional dependence.}
\label{Fig_shapes_2}
\end{figure}

As we get deeper into the occupied states   $\xi \ll 0 $, we find a remarkable change in shape of the spectral functions. This is illustrated in \figdisp{Fig_shapes} where we plot $\gamma$ of \disp{peak-3} after normalizing to unit area.
In order to have a well defined quasiparticle type peak in $\rho_G$ for $\epsilon \sim O(1)$, $Q_k$ must be small enough. This translates to the requirement of $\xi_k$ being close to the Fermi surface. {\col  By setting $Q \sim 1$ as the condition for losing a peak in the spectrum we obtain the condition
\beq
\varepsilon_0+ (1-Z_{k})\xi_k = \frac{1}{\sqrt{3}}Z_{k} \Gamma_k, \label{disappear}
\eeq
beyond which it is meaningless to talk of quasiparticles. This gives $\varepsilon_0$ as a rough characteristic scale for the disappearance of the quasiparticle peaks.}

Fig.~(\ref{Fig_shapes_2})  illustrates the change in shape somewhat differently by  normalizing all curves to unity at the peak as in \disp{peak-3}.
 The peak at $\bar{\epsilon}=0$ is sharp and quite symmetric for $Q \ll 1$, and  becomes broader and more left skewed    as $Q_k$ increases towards its maximum value $2$. Attaining the maximum value is  possible in  principle,
 requirement being $u_k=0$ or from \disp{convenient}:
\beq
\Delta_0= \frac{n^2}{4 - 2n} \ \Re e \ \Phi(\E_k).
\eeq
Unless $\Delta_0$ is very small, this condition is hard to satisfy.
If this possibility is achieved then there are several interesting consequences. Firstly we note that 
from \disp{peak-ecfl}, the magnitude of the spectral function at $u_k=0$ becomes insensitive to  
disorder and temperature etc. Its   magnitude $n^2 Z_k/( 8 \pi \Delta_0)$ that should be  useful for finding $\Delta_0$.  
 If this is approximately satisfied, then   the peak structure  loses meaning and the spectrum is essentially flat. Taking  $\bar{\epsilon}=-1$, the fall off from the peak value of unity is $80\%$, and the spectrum becomes essentially featureless.
 
\subsection{Skewness  parameter of the spectrum}
We now estimate the  skewness of the spectrum. 
The function \disp{peak-3} drops to half its peak value at two values of $\bar{\epsilon}_{L,R}$ to the left and right of the peak given by: 
\beq
\bar{\epsilon}_{L,R}= - \frac{1}{2} Q_k \mp \sqrt{Q_k+\frac{1}{4}Q_k^2},
\eeq 
For small $Q\ll1$, the (dimensionless) width of the peak is small,  it increases with $Q$ as discussed further below.
We  define a dimensionless skew parameter in terms of the energies $\epsilon^*, \epsilon_{R,L} $ (rather than the $\bar{\epsilon}$'s that are related via \disp{ebars})
\beq
\kappa(\xi_k)= \frac{ 2\epsilon^*- \epsilon_R-\epsilon_L }{\epsilon^*- \epsilon_L} = \tanh(u_k)-1+\sqrt{(2-\tanh(u_k))^2-1}, \label{skew}
\eeq
with the property that near the Fermi level when $u_k \to +\infty$ the variable  $\kappa\to 0$, and  we get a symmetric curve about the maximum. On the other hand for  deeper occupation $u_k$ decreases in magnitude towards zero, driving  $\kappa\to 0.732$ and gives a curve that is increasingly biased to the left. The asymmetry $\kappa$ grows as $O(T^2)$ at low temperatures, and it is rather large  at room temperature. 
As a rough estimate  the quasiparticle peak  is lost when  $Q_k \sim .5$ where $u_k \sim .98$ and $\kappa(\xi_k) \sim .5$. This loss of quasiparticle peak structure,  skew factor and its experimental signature  is studied in greater detail in \refdisp{gweon-ecfl}. See \figdisp{Fig_q} for a typical plot of skew parameter $\kappa$ and $Q$ versus the hole binding energy $E_k$.


  
\section{\bf Self energy of  the self energy and a  Mori type procedure. \label{secmori}}

Since the construction given above generates $\Sigma(z)$ from $G(z)$ given only the representation \disp{eq1}, we can  as well repeat the trick.  Since $\Sigma(z)$ satisfies \disp{eq-10} with a density $\rho_\Sigma(x)$ that is assumed known,  and is  analytic in the complex $z$ plane with a $1/z$ fall off, it satisfies the necessary conditions for a further decomposition. Consulting \disp{inverse},  we write down by inspection
\barray
\Sigma(z)& = &  \frac{a_{\Sigma}}{z- \bar{E}_{\Sigma} - \Sigma^{(1)}(z)} \nn \\
\Sigma^{(1)}(z) & = & \int dx \ \frac{\rho_{\Sigma^{(1)}}(x)}{z-x} \label{self-self1}
\earray
The constants $a_{\Sigma}= \int dx \ \rho_{\Sigma}(x)$ and $\bar{E}_{\Sigma}= \int dx \ x \rho_{\Sigma}(x)/a_{\Sigma}$ are known through $\rho_\Sigma(x)$. They may in practice  be conveniently determined in terms of the moments of the Greens function \cite{mori,dupuis,hess} in applications.  The spectral function is given by
\beq
\rho_{\Sigma^{(1)}}(x) =   \frac{a_\Sigma \ \rho_{\Sigma}(x)}{( \pi^2 \rho_{\Sigma}(x))^2 + \left(  \Re e \ \Sigma(x)\right)^2 }. \label{inverse3}
\eeq
 Comparing this  representation with \disp{direct}  we note the formal similarity between  $\rho_{G}(x)$  and   $\rho_{\Sigma^{(1)}}(x) $.  Thus for a Fermi liquid with momentum independent self energy, its next self energy resembles closely the spectral function $\rho_G$ especially at the Fermi energy.

We follow up briefly  on the above amusing observation, and obtain a hierarchy of self energies starting from an initial self energy given by the spectral representation \disp{eq-10} and  \disp{self-self1}. This process parallels the continued fraction  representation of analytic functions  and    seems intimately  related to the formalism developed by Mori\cite{mori,dupuis}. The latter  is expressed in the language of projection operators for Liouville operators that is less straightforward than our  simple treatment.  

In order to conform to the notation popular in the Mori formalism, we will express the variables in Laplace representation rather than the one used above with complex frequencies. 
  Let us  consider the thermal and temporal correlation function for two operators $A$ and $B$ in Schr\"odinger time and its Laplace transform:
\barray
C_{AB}(t) & = &  \int_{0}^{\beta}\ d \tau \ \langle A(t - i \tau) B(0) \rangle, \;\;\;\hat{C}_{AB}(s)  =  \int_{0}^{\infty} \ e^{- s t} \ C_{AB}(t) \ dt \
\earray
 In the standard  case we find  $A = B^{\dagger}$, where the product is real and also positive \cite{fn6}. We see that the Laplace transform function satisfies an integral representation 
\barray
\hat{C}_{AB}(s) & = & \int_{- \infty}^{\infty} \ d \nu \ \frac{\rho_{A B}(\nu)}{s - i \nu}, \;\;\mbox{with a real density given by } \nn \\
\rho_{A B}(\nu) & = &  \sum_{nm} \frac{p_{m }-p_{n}}{\varepsilon_{n}-\varepsilon_{m}} \langle n | A | m \rangle \ \langle m | B | n \rangle \ \delta(\varepsilon_{n}-\varepsilon_{m}- \nu). \label{lehmann}
\earray
This object is closely  connected with the correlation functions used in \disp{eq1} and \disp{eq2}  by using the fluctuation dissipation theorem. Following Mori we write down a  relaxation function with the normalization property $Z_{0}(0)=1$,  and its Laplace transform
\beq
Z_{0}(t) \equiv \frac{C_{AB}(t)}{C_{AB}(0)}, \;\; \hat{Z}_{0}(s)=\int_{- \infty}^{\infty} \ d \nu \ \frac{\rho_{0}(\nu)}{s - i \nu}.
 \label{iter0}
\eeq
Here  the real density $ \rho_{0}(\nu)= \frac{1}{C(0)} \rho_{AB}(\nu) $, satisfies  the normalization condition $
  \int_{- \infty}^{\infty} \ d \nu \ {\rho_{0}(\nu)}= 1.$
Using the identity
$$ \frac{1}{ 0^{+} + i (u-v)}= \pi \ \delta(u-v) - i {\cal P} \frac{1}{u-v},$$
with $ {\cal P}$ denoting the principle value, an inverse relation expressing $\rho_{0}(\nu) =\frac{1}{\pi} \Re e \ \hat{Z}_{0}(0^{+}+ i \nu)$ follows. In order to find a Dyson type  representation for $\rho_{0}$,  following \disp{eq-eom1} and \disp{eq-I} we take the ``equation of motion'' by multiplying \disp{iter0} by $s$ and write
\barray
s \hat{Z}_{0}(s) = 1 + i  \  \int_{- \infty}^{\infty} \ d \nu  \ \frac{ \nu \ \rho_{0}(\nu)}{s - i \nu}\equiv Y_{0}(s) \ \hat{Z}_{0}(s). 
\earray
 The Dyson form of self energy now emerges and we obtain:
 \beq
 \hat{Z}_{0}(s)= \frac{1}{s - i Y_{0}(s)}, \;\; \mbox{with} \;\;Y_{0}(s) =  \ \frac{\int_{- \infty}^{\infty} \ d \nu  \ \frac{ \nu \ \rho_{0}(\nu)}{s - i \nu}}{\int_{- \infty}^{\infty} \ d \nu  \ \frac{  \ \rho_{0}(\nu)}{s - i \nu}}. \label{Y0}
 \eeq
  As $s \to \infty$, the function $Y_{0}(s)$ tends to  $ \omega_{1}$, with a real frequency $\omega_1$ given by
\beq
\omega_{1}= \int_{-\infty}^{\infty}\ d \nu \ \nu \ \rho_{0}(\nu). 
\eeq
 Hence the function $(Y_{0}(s)-   \omega_{1}) $ falls off as $\frac{1}{s}$ as $ s \to \infty$. It is analytic everywhere except on the imaginary $s$  axis. It therefore has a representation
\beq
Y_{0}(s)-   \ \omega_{1}=  i \ \alpha_{1} \ \int_{- \infty}^{\infty} \ d \nu  \ \frac{  \ \rho_{1}(\nu)}{s - i \nu},
\eeq
with a real   density $ \alpha_{1} \rho_{1}(\nu)= \frac{1}{\pi} \Im m \{ Y_{0}(0^{+}+ i \nu) \}$.  With this we may write
\beq
Z_0(s)= \frac{1}{s- i \ \omega_1 + \alpha_1 \int \ d\nu \ \frac{\rho_1(\nu)}{s - i \nu} }.\label{rho2}
\eeq
 The real number $\alpha_{1}$ is found using the convention that $\rho_{1}(\nu)$ is normalized to unity. We may express $\rho_1$ solely in terms of the lower density $\rho_{0}(\nu)$ by using \disp{iter0} as:
\beq
 \ \alpha_{1} \rho_{1}(u) = \frac{ \rho_{0}(u) }{\pi^{2} \rho^{2}_{0}(u)+ \left\{ {\cal H}[\rho_{0}](u) \right\}^{2} }. \label{rho12}
\eeq
We determine $\alpha_{1}$ from Eq.~(\ref{rho12}) by integrating over $\nu$ and using the unit normalization of $\rho_{1}(u)$.  It is evident from Eq.~(\ref{rho12}) that for the physically important case of  a real and  positive initial density $\rho_0(\nu)$, the derived density $\rho_1(\nu)$ is also real positive.

This scheme is clearly generalizable to higher orders, and we simply iterate the above process. The answers may be written down by inspection as follows.
\beq
\hat{Z}_{j}(s)=  \   \int_{- \infty}^{\infty} \ d \nu \ \frac{\rho_{j}(\nu)}{s - i \nu},\;\; \mbox{with normalization:} \;\;\int_{- \infty}^{\infty} \ d \nu \ {\rho_{j}(\nu)}= 1.\label{iter1}
\eeq
These satisfy a recursion relation:
\beq
\hat{Z}_{j}(s) = \frac{1}{s - i \ \omega_{j+1} +  \alpha_{j+1} \ \hat{Z}_{j+1}(s)}, \label{iter3}
\eeq
where
\beq
\omega_{j+1}= \int_{-\infty}^{\infty} \ d \nu \ \nu \ \rho_{j}(\nu), \label{iter4}
\eeq
and $\alpha_{j+1}$ as well as $\rho_{j+1}(\nu)$ are defined through
\beq
 \ \alpha_{j+1}  \ \rho_{j+1}(u) = \frac{ \rho_{j}(u)}{\pi^{2} \ \rho^{2}_{j}(u)+  \left\{ {\cal H}[\rho_{j}](u) \right\}^{2} }. \label{iter5}
\eeq
Note that  the numbers $\alpha_j$ as well as $\omega_j$ are real, and  for all $j$, the densities $\rho_j(\nu)$ are positive provided the  the initial density $\rho_o(\nu)$ is positive. This situation arises when the initial operators $B= A^\dagger$, as mentioned above.

It is clear that \disp{rho12} is the precise analog of the relation \disp{inverse} for the Greens function. The hierarchy of equations  consisting of Eq.~(45-49) constitutes an iteration scheme  that  starts with $j=0$ correlation function in \disp{iter0}. This is a forward hierarchy in the sense that successive densities at level  $j+1$  are  expressed explicitly in terms of the earlier ones at level  $j$. In the reverse direction it is rather simpler since level $j$ is explicitly given in terms of level $j+1$ by \disp{iter3}. The use of this set of equations requires some a priori  knowledge of the behaviour of higher  order  self energies to deduce the lower ones. Standard approximations \cite{mori} consist of either truncation of the series or making a physical assumption  such as a Gaussian behaviour at some  level and then working out the lower level objects. Our object in presenting  the above  working  is merely  to point out that this iterative scheme is in essence  a rather simple application  of the   self energy concept described above, with the repeated use of \disp{inverse}. 
 
 \section{Summary and Conclusions}
A  new form of the electronic Greens function, departing widely from the Dyson form  arises in the extreme correlation  theory of the \tJ model. Motivated by its  considerable success    in explaining ARPES data  of optimally doped cuprate superconductors\cite{gweon-ecfl},   we have presented  in this paper results on the detailed structure of this Greens function and its spectral function.   An illustrative example is provided, complete with numerical results, so that the novel  line shape and its dependence on parameters is revealed. We have also presented a set of explicit results on the Mori form of the self energy that holds promise in several contexts.
  
 \section{Acknowledgements}
 This work was supported by DOE under Grant No. FG02-06ER46319.  I thank   G-H. Gweon   for helpful comments and valuable discussions. I  thank A. Dhar and P. Wolfle for  useful comments regarding the Mori formalism.


\begin{thebibliography}{36}
\bibitem{ECFL}B. S. Shastry, arXiv:1102.2858 (2011), {\em  Extremely Correlated Fermi Liquids }  Phys. Rev. Lett. {\bf 107}, 056403 (2011).
\bibitem{ECQL} B. S. Shastry, Phys. Rev. {\bf B 81}, 045121 (2010).
\bibitem{gweon-ecfl} G.-H. Gweon,  B. S. Shastry and G. D. Gu, {\em Extremely Correlated Fermi Liquid Description of Normal State ARPES in Cuprates}, arXiv:1104. (2011),  Phys. Rev. Lett. {\bf 107},  056404  (2011).

\bibitem{logan} D. E. Logan, M. P. Eastwood and M, A. Tusch, J. Phys. Cond. Mat. {\bf 10}, 2673 (1998).
\bibitem{kotliar} Z. Wang, Y. Bang, and G. Kotliar,   Phys. Rev. Lett. {\bf 67}, 2733 (1991).
\bibitem{luttinger} J. M. Luttinger and J. C. Ward, Phys. Rev {\bf 118}, 1417 (1960), J. M . Luttinger, Phys. Rev. {\bf 119}, 1153 (1960);
Phys. Rev. {\bf 121}, 942 (1961). 
\bibitem{mori} H. Mori, Prog. Theor. Phys. {\bf 33}, 423 (1965); Prog. Theor. Phys. {\bf 34}, 399 (1965).
\bibitem{dupuis} M. Dupuis, Prog. Theor. Phys. {\bf 37}, 502 (1967)
\bibitem{hess} J. J. Deisz, D. W. Hess and J. W. Serene, Phys. Rev. {\bf 55} 2089 (1997).  
\bibitem{agd} A. A. Abrikosov, L.  Gorkov and I. Dzyaloshinski, {\em Methods of Quantum Field Theory in Statistical Physics },  Prentice-Hall,
Englewood Cliffs, NJ (1963). Our $\rho_G(\xi_k,x)$ corresponds to the combination of spectral functions $A(\vec{k},x)$ and  $B(\vec{k},x)$ used here. 
\bibitem{fn1} The positive part constraint on the right of the following equation can be often omitted, we found that  it is violated very slightly $\sim 3\%$ in many cases. We omit it for simplicity in the following.
\bibitem{fn2}{Shifting $\hat{E}$ by a constant is also possible but the optimal choice is the one made here.}
\bibitem{fn3}The notation is simplified from that in \refdisp{ECFL} by calling the extremely correlated Greens function as $G$ rather than ${\cal G}$ and the  overbar in the self energy $\bar{\Phi}$ is omitted. 
\bibitem{fn4}At $T=0$ this finite value persists and  $\Re e \ G^{Peak}(\xi_k, x^*(\xi_k))$ does not diverge, so  that one might be concerned that the Luttinger Ward volume theorem is being disobeyed.  In comparison note that the standard FL Greens function behaves in a slightly different way, at any finite $T$  and $\xi_k$, at the  energy $x=\E_k$ we find both a   peak in the  spectral function $\rho_g(\xi_k,\E_k)$ and a zero of the $\Re e \ g(\xi_k,\E_k)$, whereas at $T=0$, we find  a    delta peak in the  spectral function $\rho_g(\xi_k,\E_k)$ and a  pole of the $\Re e \ g(\xi_k,\E_k)$.  However we see that   the  FL divergence of  the real part of $G$ of
  a typical Fermi liquid does occur, but displaced by a very small energy  scale    $O(T^4)$. The peak positions are displayed in \figdisp{Fig_a} at a high enough temperature so that the features are distinguishable.
  \bibitem{mdc}   The process of scanning $\xi_k$ used here differs slightly from the true MDC's, where one scans the wave vector $\vec{k}$ rather than the energy $\xi_k$, but is more convenient here. 
  \bibitem{fn6} The spectral function $\rho_{A A^\dagger}(x)$ defined below in \disp{lehmann},  is  positive as well.  However that condition can be relaxed  and we can do with  less  provided that the spectral function in \disp{lehmann} is real (rather than positive). For this to happen, we may    allow for $A \neq B^{\dagger}$, but  in this case assume that {\em both}  the matrix elements $\langle n | A | m \rangle $ and $ \langle m | B | n \rangle $ are {\em  real numbers} or {\em imaginary numbers} so that the product is real. These conditions  correspond to both operators  operators $A, B$  being  Hermitean (or anti Hermitean), and   the absence of magnetic fields so that the wave functions may be chosen to be real.
Thus we will  assume the reality of the product of the matrix elements.
\end{thebibliography}
\end{document}